\documentclass[conference]{IEEEtran}
\IEEEoverridecommandlockouts
\usepackage{cite}
\usepackage{amsmath,amssymb,amsfonts}
\usepackage{algorithmic}
\usepackage{graphicx}
\usepackage{textcomp}
\usepackage{xspace}
\def\BibTeX{{\rm B\kern-.05em{\sc i\kern-.025em b}\kern-.08em
    T\kern-.1667em\lower.7ex\hbox{E}\kern-.125emX}}

\usepackage{xcolor,colortbl}
\usepackage[ruled,vlined]{algorithm2e}
\SetKwInput{KwInput}{Input}
\SetKwInput{KwOutput}{Output}
\usepackage[most]{tcolorbox}
\usepackage{tabularx}
\usepackage{newfloat}
\usepackage{listings}
\usepackage{soul}
\usepackage{color} 
\usepackage{caption} 
\usepackage{multirow}
\usepackage{url}
\usepackage[breaklinks=true]{hyperref}
\usepackage{enumitem}
\usepackage{dblfloatfix}

\newtcolorbox{boxEnv_rq}{
    center,
    left=0 mm,
    top = 0.25 mm,
    right = 0mm,
    bottom =0.25 mm,
    colframe=gray!90!black,
    colback=black!5!white, 
    boxrule=0.5pt,
    title = Research Questions (RQ),
}

\newtcolorbox{boxEnv_ps}{
    center,
    left=0 mm,
    top = 0.25 mm,
    right = 0mm,
    bottom =0.25 mm,
    colframe=gray!90!black,
    colback=black!5!white, 
    boxrule=0.5pt,
    title = Problem Statement,
}

\newtcolorbox[auto counter]{appbox}[2][]{
    enhanced, breakable,
    attach boxed title to top center={yshift=-3mm,yshifttext=-1mm},
    colback=white, colframe=gray!75!black, colbacktitle=gray!80!black,
    title={Box~\thetcbcounter: #2},
    boxed title style={size=small,colframe=gray!90!black},
    left=0.5mm, right=0.5mm, boxrule=0.75pt, fontupper=\small, #1
}

\definecolor{softgreen}{RGB}{198,239,206}
\definecolor{softblue}{RGB}{197,217,241}

\newcommand{\falcon}{FALCON\xspace}
\newcommand{\signature}{\mathcal{S}}
\newcommand{\behavior}{\mathcal{B}}
\newcommand{\cti}{\mathcal{C}}
\newcommand{\idsrule}{\mathcal{R}}
\newcommand{\instruction}{\mathcal{I}}
\newcommand{\feedback}{\mathcal{F}}
\newcommand{\threshold}{\mathcal{T}}

\begin{document}

\title{\falcon: Transforming Cyber Threat Intelligence into Deployable IDS Rules with Self-Reflection\\}


\author{
\IEEEauthorblockN{
Shaswata Mitra\IEEEauthorrefmark{1}, 
Subash Neupane\IEEEauthorrefmark{2}, 
Martin Duclos\IEEEauthorrefmark{3},
Sudip Mittal\IEEEauthorrefmark{4},\\
Aritran Piplai\IEEEauthorrefmark{5},
Md Rayhanur Rahman\IEEEauthorrefmark{6},
Edward Zieglar\IEEEauthorrefmark{7}
Shahram Rahimi\IEEEauthorrefmark{8},
}
\IEEEauthorrefmark{2}Meharry Medical College -- \{\IEEEauthorrefmark{2}subash.neupane@mmc.edu\}\\
\IEEEauthorrefmark{5}The University of Texas at El Paso -- \IEEEauthorrefmark{5}apiplai@utep.edu\\
\IEEEauthorrefmark{1}\IEEEauthorrefmark{3}\IEEEauthorrefmark{4}\IEEEauthorrefmark{6}\IEEEauthorrefmark{8}The University of Alabama \\ \{\IEEEauthorrefmark{1}smitra3, \IEEEauthorrefmark{3} mjduclos, \IEEEauthorrefmark{4}sudip.mittal, \IEEEauthorrefmark{6}mdrayhanur.rahman, \IEEEauthorrefmark{8}shahram.rahimi\}@ua.edu\\
\IEEEauthorrefmark{7}National Security Agency -- \IEEEauthorrefmark{7}evziegl@uwe.nsa.gov \\
}

\maketitle

\begin{abstract}

    Signature-based Intrusion Detection Systems (IDS) detect malicious activity by matching network or host events against predefined rules. Security analysts manually develop these rules from Cyber Threat Intelligence (CTI). As threats evolve, this manual pipeline faces two bottlenecks. Before authoring a new rule, an analyst must reconcile the incoming CTI with the existing rule base and determine whether to create, update, or retire one. This process is challenging due to the representational differences between the CTI and Rule formats. This gap limits the effectiveness of keyword- and embedding-based search, making rule reconciliation cognitively demanding and, in turn, contributing to ``rule bloat''. Second, automated verification of a new rule is inherently difficult as zero-day threats lack ground truth from simulated testing. Hence, standard metrics cannot prove that a rule semantically adheres to the CTI, and the use of LLMs leads to non-deterministic behavior. To address these challenges, we introduce \falcon\footnote{\label{github}Code \& Dataset: \url{github.com/shaswata09/falcon}}, an agentic framework for CTI-grounded rule retrieval, generation, and validation. At its core, a novel CTI-Rule semantic scorer\footnote{\label{hf}Models: \url{huggingface.co/collections/shaswatamitra/falcon}}, quantifies the functional alignment between a CTI and a rule; the same signal drives a retriever that surfaces relevant deployed rules and a ground-truth-free validator that scores generated ones. Around it, a generation pipeline produces deployable rules from CTI in real time and refines them through self-reflective syntactic, semantic, and performance validators. Across network (Snort) and host-based (YARA) platforms on a purpose-built CTI-Rule dataset, \falcon attains a mean relevance of $\approx$0.72, with 84\% inter-rater agreement among cybersecurity analysts, underscoring the promise of real-time security automation.

\end{abstract}


\section{Introduction}

    Every year, approximately 7 trillion intrusion attempts are made globally, with over 90\% of breaches exploiting \textit{known, unpatched vulnerabilities} ~\cite {cybersecurity_stat}. Signature-based Intrusion Detection Systems (IDSs) counter such threats by continuously monitoring networks and hosts and matching activity against predefined rules that encode attack signatures and behaviors. Security Operations Center (SOC) analysts are responsible for developing these rules through a structured rule generation process. This process involves several phases, including retrieving relevant deployed rules, analyzing threat behavior, generating rules, and assessing their compatibility with the existing rule base. This process is iterative, where findings are accumulated in Cyber Threat Intelligence (CTI) to enable optimal rule generation. This IDS rule-generation process from CTI is consistent across network and host media and mainly relies on analyst expertise to ensure optimal rule generation.

    \begin{figure}[!ht]
      \centering \includegraphics[width=0.48\textwidth]{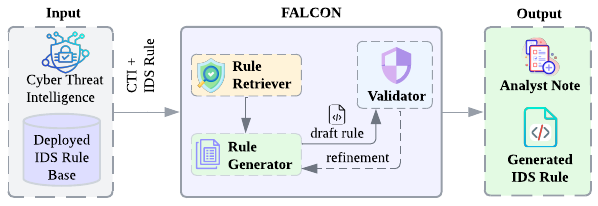}
      \caption{TLDR; \falcon accepts CTI and deployed IDS rules, produces an IDS rule and analyst note for explainability. The rule is generated through orchestration of a Rule Retriever, LLM based Generator, and a deterministic Validator.} 
      \label{fig:tldr}
    \end{figure}

    This manual, cognitively heavy pipeline struggles to keep pace with a fast-evolving threat landscape. As attackers vary their Tactics, Techniques, and Procedures (TTPs), new signatures, behaviors, and matching rules proliferate. In contrast, existing rules decay~\cite{chinwe2025adversarial}, and the daily volume of CTI, including sandbox traces, behavioral logs, and IoCs, makes \textit{manual analysis burdensome and cognitively demanding}~\cite{wack2002guidelines}. Two problems make this especially hard to automate. \textbf{First}, a new rule cannot be generated in isolation: the analyst must search through the deployed rule base for functionally relevant rules and decide whether to create, update, or deprecate one. Skipping this reconciliation inflates the rule base (`\textit{rule-bloat}'), straining IDS engine performance and functionality, yet performing it demands expertise and assumes \textit{relevant rules are retrieved beforehand to enable context-aware generation}. The representational gap between natural language or structured CTI and formal rule syntax causes keyword and embedding-similarity search to fail at exactly this retrieval. \textbf{Second}, \textit{without ground truth, it is difficult to deterministically determine whether an LLM-generated rule meets the functional requirements of the CTI}~\cite{jin2024simllm}, as it is challenging to simulate and test zero-day threats. Hence, any delay or error in deployment leaves systems exposed without any defense.

    IDS rules are also platform-dependent. Network rules such as Snort detect and block malicious traffic, whereas host rules such as YARA flag system-level anomalies, for example memory or registry changes, by matching textual or binary patterns. Covering both demands rules tailored to each platform's capabilities, compounding the manual effort~\cite{bridges2019survey}. Accordingly, we ask the following research questions (RQs):

    \begin{boxEnv_rq}
        \begin{itemize}
            \item \textbf{RQ-1:} To what extent can LLMs translate heterogeneous CTI into deployable IDS rules that are functionally aligned across detection mediums?

            \item \textbf{RQ-2:} Given the representational gap between heterogeneous CTI and formal rule syntax, and the absence of ground truth at test-time, how reliably can the functional alignment between a CTI and an IDS rule be quantified?

            \item \textbf{RQ-3:} Can structured feedback enable LLM to iteratively refine a sub-optimal rule with measurable quality gains over one-shot generation?

        \end{itemize}
    \end{boxEnv_rq}

    To address these challenges, we present \falcon (Feedback-driven ALignment framework for Cti-to-rule generatiON), a self-reflective agentic framework for IDS rule generation and improvement. \falcon builds on AI-based signature and behavior extraction from threat analysis~\cite{guan2024logllm, mitra2024localintel, rahman2024chronocti} and uses agentic LLMs to automate the full pipeline: given a CTI input of behavioral descriptions, signatures, or IoCs, it autonomously produces deployable rules for the target environment [Fig.~\ref{fig:tldr}]. At its core, a novel \textit{CTI-to-Rule Semantic Scorer} quantifies the functional alignment between a CTI and a rule. The same signal drives both a \textit{retriever} that surfaces relevant deployed rules and a \textit{deterministic validation pipeline} that checks each generated rule for syntactic correctness, semantic alignment, and performance for self-reflection and explainability. This couples relevant rule reuse with an approximation of test-time quality assessment, enabling adaptive rule generation while curbing `rule bloat'. 
    
    Our contributions include:
    \begin{itemize}
        \item A \textit{self-reflective framework} that autonomously transforms heterogeneous CTI into deployable IDS rules across both network (Snort) and host (YARA) mediums (RQ-1).

        \item A \textit{CTI-Rule semantic alignment model} that quantifies the functional correspondence between a CTI and a formal rule without ground truth, and show that a single such model serves a dual role, retrieving relevant deployed rules and validating generated ones (RQ-2).

        \item A \textit{self-reflective validation pipeline} that diagnoses a candidate rule along syntactic, semantic, and performance dimensions and returns feedback, enabling an LLM agent to iteratively refine sub-optimal rules with measurable gains over one-shot generation (RQ-3).

        \item A \textit{comprehensive dataset} that mimics real-world scenarios, pairing ground-truth and relevant-but-outdated IDS rules with their corresponding CTIs, enabling both quantitative evaluation and qualitative expert validation of \falcon (RQ-1, 2, 3). 
    \end{itemize}

    The remainder of the paper covers preliminaries (Section~\ref{section:background}), problem formulation (Section~\ref{section:problem_formulation}), the \falcon framework (Section~\ref{section:aidrag_framework}), experiments and evaluation (Section~\ref{section:experiment}), and concluding remarks with future directions (Section~\ref{section:conclusion}).

\section{Preliminaries and Motivation}\label{section:background}

    \subsection{From Hand-Crafted to LLM-Generated Rules}
        Automated signature generation predates LLMs by nearly two decades. Classical NIDS methods derive signatures from honeypots or captured traffic~\cite{newsome2005polygraph, kreibich2004honeycomb}, and HIDS methods build YARA rules from malware corpora using $n$-grams, biclustering, or learned scorers~\cite{raff2020autoyara, li2023packgenome, naik2020evaluating, IEEE10693305}. These operate on raw binaries or pre-captured traffic and offer no mechanism to reconcile new rules against a deployed rule base.

        LLMs have shifted the input from raw artifacts toward analyst-facing text, and a fast-growing line now synthesizes detection rules from cloud OSCTI~\cite{schwartz2024llmcloudhunter}, honeypot PCAPs~\cite{balasubramanian2024hex2sign}, package metadata~\cite{zhang2025rulellm}, proof-of-concept exploits~\cite{lian2025rulemaster}, ICS network flows~\cite{moreno2025leveraging}, or live attack traffic in a cyber range~\cite{du2025harnessing}, with recent multi-agent variants that additionally repair rules against an existing ruleset to limit redundancy~\cite{li2025gridai}; a parallel line curates deployed SIEM rule sets post hoc~\cite{shukla2025rulegenie}. Across this work, two properties recur: the input is typically an observable artifact rather than the heterogeneous linguistic CTI analysts actually author, and each system targets a single platform.

    \subsection{Open Gaps}
        \textbf{\textit{Lack of deterministic assessment:}} Judging whether a generated rule is functionally faithful to its CTI is mostly qualitative and relies on analyst expertise and post-hoc trials. In contrast, quantitative evaluations require information unavailable for novel threats. Deployment-based evaluation replays labeled attack traffic in a live engine~\cite{li2025gridai, moreno2025leveraging, du2025harnessing}, presuming the threat is already observable and thus cannot judge a rule synthesized from scratch. Reference-based evaluation compares against a held-out human rule using lexically incompatible metrics such as ROUGE or BLEU~\cite{wang2025rulepilot, bertiger2025evaluating}. Furthermore, LLM-as-judge~\cite{guerdan2026validating} scoring offers no calibrated, reproducible signal. The field thus lacks a way to quantify, without ground truth, how well a generated rule captures its CTI.

        \textbf{\textit{No paired CTI-to-rule corpus exists:}} Detection rules (e.g., Snort, YARA) and CTI reports are each abundant but siloed, and the CTI-narrative-to-rule mapping that analysts produce by hand or automatically is not recorded in any public dataset. The artifacts prior to the generator's release are bound to a different input or a single platform: RuleLLM~\cite{zhang2025rulellm} open-sources tooling over package code, LLMCloudHunter~\cite{schwartz2024llmcloudhunter} maps a small set of cloud reports to Sigma, and Hex2Sign~\cite{balasubramanian2024hex2sign} derives Suricata rules from honeypot PCAPs. At the same time, substantive real-world corpora remain private~\cite{li2025gridai}. None pairs analyst-authored CTI with deployable rules across both NIDS and HIDS, and standard keyword or embedding similarity does not bridge the representational gap between prose CTI and formal rule syntax.

        The above-mentioned gaps shape \falcon. We release\footref{hf}, to our knowledge, the first paired CTI-rule dataset spanning Snort and YARA, augmented with relevant-but-outdated rules for retrieval, and a learned CTI-Rule alignment scorer that supplies the missing fidelity signal for both retrieval and validation. We formalize the problem in Section~\ref{section:problem_formulation}.

\section{Problem Formulation}\label{section:problem_formulation}

    \begin{table}[h]
    \caption{Description of Notation.}
        \renewcommand{\arraystretch}{1.0}
        \begin{center}
        \scriptsize
        \begin{tabularx}{0.45\textwidth} {
          c|>{\raggedright\arraybackslash}X }
         \hline
         \rowcolor{lightgray}
         \textbf{Notation} & \textbf{Description} \\
        \hline
        \hline
        $ \emptyset $ & Null Set \\
        $ \instruction $ & Generation Instruction (Constant) \\
        $ \threshold $ & Validation Threshold (Constant) \\
        $ \mathcal{E}_c,\ \mathcal{E}_r $ & CTI and Rule Embedding Encoders \\
        $ \Phi(\cti, \idsrule) $ & CTI-Rule Alignment Scorer \\
        $ \{\cti_i \mid \cti_i \in \cti \}$ & Set of Cyber Threat Intelligence \\
        $ \{\feedback_i \mid \feedback_i \in \feedback\}$ & Set of Validation Feedback \\
        $ \{\feedback^s, \feedback^f, \feedback^p, \feedback^a\}$ & Syntax, Semantic, Performance, Analyst Feedback \\
        $\{ \idsrule^e_i \mid \idsrule^e_i \in \idsrule \}$ & Existing and Deployed IDS Rule Set \\
        $\{ \idsrule_i \mid \idsrule_i \in \idsrule \}$ & Generated IDS Rule Set \\
        $ \epsilon \in [0, 1] $ & Semantic Alignment Score ($\cti$ vs $\idsrule$) \\
        \hline
        \end{tabularx}
        \end{center}
    \end{table}

    Threat signatures $\signature = \{\signature_0, \signature_1, \dots, \signature_n \}$ are identifiable static indicators of malicious activity, such as IP addresses, hashes, or other Indicators of Compromise (IoCs). Threat behaviors $\behavior = \{\behavior_0, \behavior_1, \dots, \behavior_n \}$ are dynamic patterns, such as protocol usage, file types, or signature combinations that characterize malicious operations. A CTI instance $\cti_i$ encapsulates structured or unstructured knowledge containing signatures, behaviors, or both, and the corpus is $\cti = \{\cti_0, \cti_1, \dots, \cti_n \}$.

    \textbf{Hypothesis:} We assume each CTI contains the relevant signatures and behaviors [$\cti_i \cap (\signature_i \cup \behavior_i) \neq \emptyset $], that is, it carries the information needed for rule generation. Without this condition, \falcon cannot recover $\signature_i$ and $\behavior_i$ from $\cti_i$ to contextualize against the deployed rule base $\idsrule^e$.

    \begin{boxEnv_ps}
        Given an incoming CTI $\cti_i$ and a deployed rule base $\idsrule^e$, the task is to produce a relevant, deployable rule $\idsrule_i$ that reuses or reconciles existing rules. Let $f$ map a CTI to a rule whose output corresponds to the threat:
        \begin{equation}\label{eq:task}
            \idsrule_i = f(\cti_i, \idsrule^e) \qquad \text{where}, \idsrule_i \cap \cti_i \neq \emptyset .
        \end{equation}
    \end{boxEnv_ps}

    We decompose $f$ into three coupled sub-problems, aligned with our RQ-1 to RQ-3.\\

    \textit{(i) Functional alignment without ground truth (RQ-2).} Since no reference rule exists at test time, we require a deterministic, learned quantitative scorer $\Phi$ that maps a CTI-rule pair to a bounded alignment score,
    \begin{equation}\label{eq:scorer}
        \Phi:\ (\cti_i, \idsrule)\ \mapsto\ \epsilon \in [0,1],
    \end{equation}
    with $\epsilon \to 1$ for functionally aligned pairs. $\Phi$ is the only quality signal available for downstream evaluation. Hence, shared by the next two sub-problems.\\

    \textit{(ii) Context-aware retrieval and generation (RQ-1).} Using $\Phi$, \falcon retrieves the relevant deployed rules and records the strongest existing match,
    \begin{equation}\label{eq:retrieval}
        \idsrule^e_i = \underset{\idsrule \in \idsrule^e}{\mathrm{Top}\text{-}k}\ \Phi(\cti_i, \idsrule), \qquad \epsilon_i^{\max} = \max_{\idsrule \in \idsrule^e_i} \Phi(\cti_i, \idsrule),
    \end{equation}
    then generates a candidate conditioned on the CTI and this retrieved context, $\idsrule_i = f(\cti_i, \idsrule^e_i)$.

    \textit{(iii) Self-reflective refinement (RQ-3).} The candidate is validated into structured feedback $\feedback_i = \feedback^s_i \cup \feedback^f_i \cup \feedback^p_i$, whose semantic term $\feedback^f_i$ is derived from $\Phi(\cti_i, \idsrule_i)$. \falcon then refines iteratively,
    \begin{equation}\label{eq:accept}
        \idsrule_i^{(t+1)} = f\!\left(\cti_i, \idsrule^e_i, \feedback_i^{(t)}\right) \ \ \text{until} \ \ \feedback_i \geq \threshold \ \wedge\ \Phi(\cti_i, \idsrule_i) \geq \epsilon_i^{\max},
    \end{equation}
    so that the accepted rule passes every validation phase and is more aligned with $\cti_i$ than the best retrieved rules.

    Together, $\Phi$ links retrieval and validation through a single ground-truth-free signal, while $f$ and the feedback loop realize generation and refinement. Section~\ref{section:aidrag_framework} instantiates each component.

    \begin{figure*}[!ht]
      \centering \includegraphics[width=0.85\textwidth]{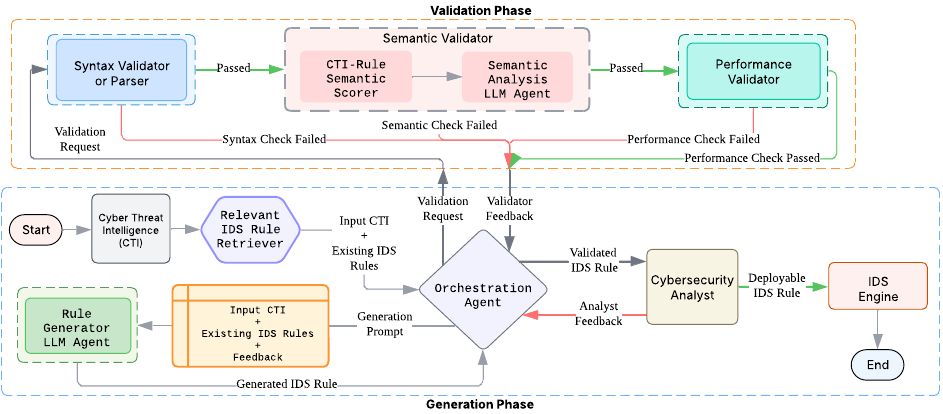}
      \caption{\falcon architecture, divided into Generation and Validation phases. The Rule Generator produces a candidate IDS rule from the CTI and retrieved rules, which is then validated for syntax, semantics, and performance. The Orchestration Agent routes validation feedback and triggers regeneration, and validated rules are reviewed by an analyst before deployment.}
      \label{fig:data_flow_diagram}
    \end{figure*}

\section{\falcon Framework}\label{section:aidrag_framework}
    We first present the overall solution approach, then describe each modules [Fig.~\ref{fig:data_flow_diagram}], and finally walk through an end-to-end use case that generates an IDS rule ($\idsrule_i$) from a CTI ($\cti_i$).

    \subsection{Solution Approach}\label{subsection:solution_approach}

        To test \textbf{RQ-1}, \falcon [Fig.~\ref{fig:data_flow_diagram}] operates in two phases for autonomous IDS rule ``generation'' and ``validation''.

        \begin{algorithm}[!ht]
            \small
            \caption{Orchestration Agent Pseudo-code} \label{alg:solution_approach}
            \KwInput{Cyber Threat Intelligence ($\cti_i)$)}
            \KwOutput{Relevant IDS Rule ($\idsrule_i \leftarrow f(\cti_i, \ \idsrule^e_i)$)}
            \
            \\
            \textsc{\textbf{Generation Phase:}} \\
            $\idsrule^e_i = \underset{\idsrule \in \idsrule^e}{\mathrm{Top}\text{-}k}\ \Phi(\cti_i, \idsrule) \leftarrow find\_relevant\_rules (\cti_i, \idsrule)$ \\
            $\epsilon_i^{max} = \max_{\idsrule \in \idsrule^e_i} \Phi(\cti_i, \idsrule)$ \\
            $\idsrule_i \leftarrow generate\_rule (\cti_i, \ \idsrule^e_i, \ \emptyset)$ \\
            $\feedback_{i} \leftarrow execute\_validation(\idsrule_i, \ \idsrule^e_i, \ \epsilon_i^{max})$\\
            \While{$\feedback_{i} < \threshold$}{
                $\idsrule_{i} \leftarrow \idsrule_{i+1} \leftarrow generate\_rule (\cti_i, \ \idsrule^e_i, \ \feedback_{i})$ \\
                $\feedback_{i} \leftarrow \feedback_{i+1} \leftarrow execute\_validation(\cti_i, \ \idsrule_{i}, \ \epsilon_i^{max})$\\
            }
            return $\idsrule_{i}$\\
            \
            \\
            \textsc{\textbf{Validation Phase [$execute\_validation$]:}}\\
            $\feedback_i \leftarrow \ \feedback_i^s \ \cup \ \feedback_i^f \ \cup \ \feedback_i^p \  \leftarrow \ \emptyset $ \\
            $\feedback_i \leftarrow \feedback_i^s \ \cup \ syntactic\_validator(\idsrule_i)$\\
            \eIf{$\feedback_i^s \ == \ \texttt{False}$}{
                return $\feedback_i$
                }{
                    $\feedback_i \leftarrow \feedback_i^f \ \cup \ semantic\_validator(\cti_i, \ \idsrule_i, \ \epsilon_i^{max})$\\
                    \eIf{$\feedback_i^f < \epsilon_i^{max} \ \ \text{or} \ \ \feedback_i^f < \threshold^f$}{
                        return $\feedback_i$
                    }{
                        $\feedback_i \leftarrow \feedback_i^p \ \cup \ performance\_validator(\idsrule_i, \ \idsrule^e_i)$\\
                    }
                }
            return $\feedback_i$
        \end{algorithm}

        \begin{itemize}
          \item[\labelitemii] The \textbf{Generation Phase} begins with an input CTI ($\cti_i$). \falcon first retrieves the deployed rules ($\idsrule^e_i$) relevant to $\cti_i$ to contextualize generation, and records their maximum semantic score ($\epsilon_i^{max}$) for later validation. The \textit{Rule Generator LLM Agent} is then invoked with the generation instruction ($\instruction$), $\cti_i$, and $\idsrule^e_i$ to produce an initial candidate ($\idsrule_i$), where $\instruction$ encodes the extraction methods and generation guidelines the agent must follow. The candidate passes to a chain of serial validators; whenever it fails the validation threshold ($\threshold$), the returned feedback ($\feedback_i$) drives iterative regeneration ($\idsrule_{i+1}$) until all criteria are met~\cite{sudarshan2024agentic}. A SOC analyst finally reviews the validated rule, approves it for deployment, and retires outdated ones; the analyst may also supply structured or unstructured feedback ($\feedback^a$) to trigger regeneration, sustaining a real-time security posture.

          \item[\labelitemii] The \textbf{Validation Phase} evaluates $\idsrule_i$ through serial \textbf{syntactic}, \textbf{semantic}, and \textbf{performance} checks. The rule is first parsed for structural correctness; if valid, semantic analysis assesses its logical consistency and functional alignment with $\cti_i$, requiring a score higher than the best retrieved match; upon passing, performance validation confirms operational effectiveness, including reconciliation with existing rules ($\idsrule^e_i$), runtime efficiency, and reliability. Failure at any stage returns immediate feedback for targeted refinement in the next iteration.
        \end{itemize}

    \subsection{\falcon System Modules}\label{subsection:system_modules}


        \subsubsection{Cyber Threat Intelligence (CTI) or $\cti$}\label{subsubsection:cti}
            CTI is the primary input to \falcon. Extracted from threat reports or logs, it comprises signatures ($\signature$) and behaviors ($\behavior$), such as IP addresses, hashes, IoCs, and protocols, forming the semantic basis for rule generation.

        \subsubsection{Relevant IDS Rule Retriever}\label{subsubsection:ids_retriever}
            The retriever identifies deployed rules ($\idsrule^e_i$) relevant to $\cti_i$, supplying context that lets the \textit{Rule Generator LLM Agent} decide whether to author a new rule or update an existing one. Because $\cti$ and $\idsrule$ have dissimilar representations, traditional embedding cosine similarity [$\text{sim}\!\left(\mathcal{E}(\cti),\, \mathcal{E}(\idsrule^e)\right) \approx \epsilon \ | \ \epsilon \to 0$] and keyword search (TF-IDF) both fail to capture functional relevance. A dedicated retriever is therefore required to draw $\cti$ and $\idsrule$ embeddings into a shared space for context-aware retrieval.

        \subsubsection{Rule Generator LLM Agent}\label{subsubsection:rule_generator}
            An LLM agent that generates $\idsrule_i$ from $\cti_i$, $\idsrule^e_i$, $\instruction$, and $\feedback_i$, interpreting feedback to iteratively refine the rule in the context of $\cti_i$.

        \subsubsection{Generated IDS Rule or $\idsrule_i$}\label{subsubsection:controller}
            The output of \falcon: an actionable rule that transforms $\cti_i$ while accounting for existing rules ($\idsrule^e_i$), refined across iterations until all validation passes.

        \subsubsection{Validator Feedback or $\feedback$}\label{subsubsection:feedback}
            The report produced by the syntax, semantic, and performance validators or the analyst. It serves two roles, evaluating $\idsrule_i$ against $\threshold$ and guiding subsequent generations, by pairing a numeric value for thresholding with descriptive, unstructured guidance for the \textit{Rule Generator LLM Agent} and analyst.

        \subsubsection{Syntax Validator}\label{subsubsection:syntax_validator}
            Verifies that $\idsrule_i$ conforms to the target engine's syntax, checking components such as headers, conditions, and options. On error, it returns a negative binary (False) signal with the parser's error description for regeneration.

        \subsubsection{Semantic Validator}\label{subsubsection:semantic_validator}
            The Semantic Validator assesses whether $\idsrule_i$ logically captures $\cti_i$, e.g., protocols, payload signatures, and behavioral patterns consistent with the detection objective. The representational gap between heterogeneous language-formatted CTI and formal rule syntax renders embedding-based similarity or existing graph-matching approaches inapplicable. At the same time, direct LLM judgment is unreliable in first-shot inference due to hallucination~\cite{liu2024exploring} and degraded long-context performance~\cite{liu2023your}. Unlike code, a zero-day threat cannot be simulated with mock test cases to verify functional alignment. We therefore required a shared latent space that projects the logical correspondence across representations, that is, in our case, between $\cti$ and $\idsrule$. Hence, a bi/dual encoder is trained to embed both into a cross-modal shared projection latent space for similarity computation~\cite{radford2021learning}. We further detail our model in  Section~\ref{subsection:semantic_scorer}.

            The resulting score feeds a \textit{Semantic Analysis LLM agent} that, guided by this quantified signal and structured prompts, predicts logical inconsistencies in $\idsrule_i$; such context-quantified guidance has been shown to improve LLM reasoning and reliability~\cite{sudarshan2024agentic}. For efficiency, we fold these inconsistencies into the next regeneration run rather than emitting them separately, since regeneration occurs only when the candidate is less aligned than the retrieved rules [$\epsilon_i^{\idsrule_i}<\epsilon_i^{max}$] or below threshold [$\epsilon_i^{\idsrule_i}<\threshold $]. Because the improvement-feedback tokens precede the next rule's tokens, they set the correct generation trajectory for $\idsrule_{i+1}$ without sacrificing accuracy, echoing chain-of-thought prompting~\cite{yao2023react}.

        \subsubsection{Performance Validator}\label{subsubsection:performance_validator}
            The Performance Validator assesses the operational efficiency of $\idsrule_i$ for a production environment. For example, matching many signatures via sequential \texttt{if-else} logic is functional but inferior to a shared regular expression. Such checks can use traditional algorithms, such as DAG analysis~\cite{abbes2010efficient, wang2023dag}, or separate LLM inference; LLM agents have been adapted to evaluate optimization criteria~\cite{gao2024search} accurately. For this work, we adopt traditional validators with a lenient threshold to assess execution speed, resource utilization, and rule reusability, with detection coverage; rules with excessive latency, redundant logic, or unnecessary additions are returned with negative feedback ($\feedback^p$). Since some existing rules can address a zero-day with minimal change, each retrieved rule $\idsrule_i^e$ is assigned a \texttt{CREATE}, \texttt{UPDATE}, or \texttt{DELETE} label. Given the domain's sensitivity, where an erroneous deletion or addition is costly, these labels are surfaced as an \textit{Analyst Note} for SOC approval rather than applied automatically, ensuring only high-performing, compatible rules advance with accountability.

        \begin{figure*}[!ht]
          \centering \includegraphics[width=0.8\textwidth]{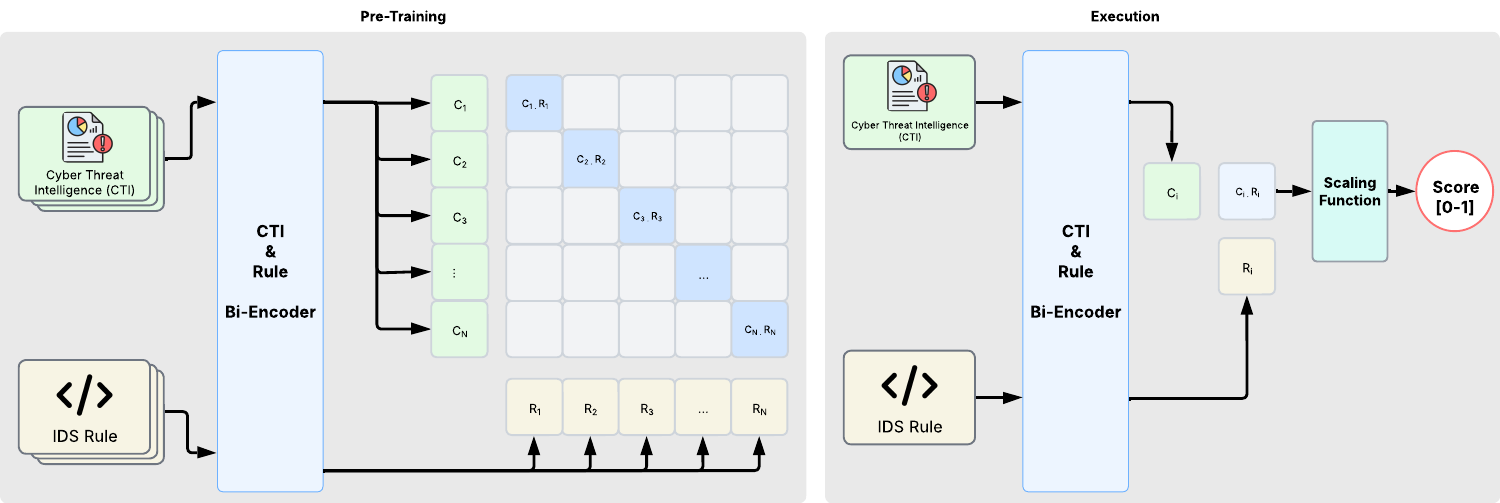}
          \caption{Semantic Scorer training and execution diagram. A bi-encoder model quantifies semantic similarity between $\cti$ and $\idsrule$ on a [0--1] scale. Each $\cti_i$ and $\idsrule_i$ is encoded independently, and cosine similarity populates the matrix. During contrastive pre-training, correct $(\cti_i, \idsrule_i)$ pairs (diagonal entries) are optimized for the highest softmax score to capture logical alignment. At execution, the trained model scores a new candidate rule $\idsrule_i$ for semantic consistency w.r.t. the input CTI $\cti_i$.}
          \label{fig:semantic_scorer_model}
        \end{figure*}

    \subsection{\falcon CTI-Rule Semantic Scorer Model}\label{subsection:semantic_scorer}

        To answer \textbf{RQ-2}, we instantiate the scorer $\Phi$ of Eq.~\eqref{eq:scorer}. Code-similarity models (GraphCodeBERT~\cite{guo2020graphcodebert}, CodeBERT~\cite{feng2020codebert}) capture structural equivalence but ignore detection intent and rule syntax, and lexical metrics (ROUGE~\cite{lin2004rouge}, BLEU~\cite{papineni2002bleu}) miss the CTI–rule relationship given their wide gap in length and abstraction. We instead encode each input independently with $\mathcal{E}_c,\mathcal{E}_r$, giving $\mathbf{z}_{\cti_i}=\mathcal{E}_c(\cti_i)$, $\mathbf{z}_{\idsrule}=\mathcal{E}_r(\idsrule)\in\mathbb{R}^{768}$ [Fig.~\ref{fig:semantic_scorer_model}], and score by a calibrated cosine,
        \begin{equation}\label{eq:phi}
            \Phi(\cti_i,\idsrule) = \sigma\!\big(\cos(\mathbf{z}_{\cti_i},\mathbf{z}_{\idsrule})\big) = \epsilon \in [0,1].
        \end{equation}
        Independent encoding keeps $\Phi$ cheap over the full base $\idsrule^e$ at retrieval and each refinement step. We evaluate both bi- and dual-encoder configurations under contrastive fine-tuning (below); both perform well in practice [see repository Appendix\footref{github}].

        \textbf{Latent regularization:} A deterministic autoencoder is inadequate: its reconstruction objective rewards modality-specific surface features, the very cues separating $\cti$ from $\idsrule$, and its point estimates leave a sparse, discontinuous latent where cosine is unreliable and contrastive training collapses onto an anisotropic cone. We thus use a VAE-style stochastic encoder $q(\mathbf{z}\mid x)=\mathcal{N}\!\big(\boldsymbol{\mu}(x),\mathrm{diag}(\boldsymbol{\sigma}^2(x))\big)$, $x\in\{\cti_i,\idsrule_i\}$, reparameterized $\mathbf{z}=\boldsymbol{\mu}(x)+\boldsymbol{\sigma}(x)\odot\boldsymbol{\eta}$, $\boldsymbol{\eta}\sim\mathcal{N}(0,I)$, regularized to an isotropic prior over the fixed $768$-dimensional latent,
        \begin{equation}\label{eq:kl}
            \mathcal{L}_{\text{KL}} = \mathbb{E}_{x}\,D_{KL}\!\big(q(\mathbf{z}\mid x)\,\|\,\mathcal{N}(0,I)\big)
        \end{equation}
        This fills latent holes (smooth, calibrated cosine geometry), injects noise that resists collapse, and aligns the distributions $q(\mathbf{z}\mid\cti_i)$ and $q(\mathbf{z}\mid\idsrule_i)$ as overlapping clusters rather than points, robust under limited data, yielding $\mathcal{Z}_{\cti}\overset{\text{func}}{\leftrightarrow}\mathcal{Z}_{\idsrule}$.

\begin{figure*}[!ht]
        \centering \includegraphics[width=0.9\textwidth]{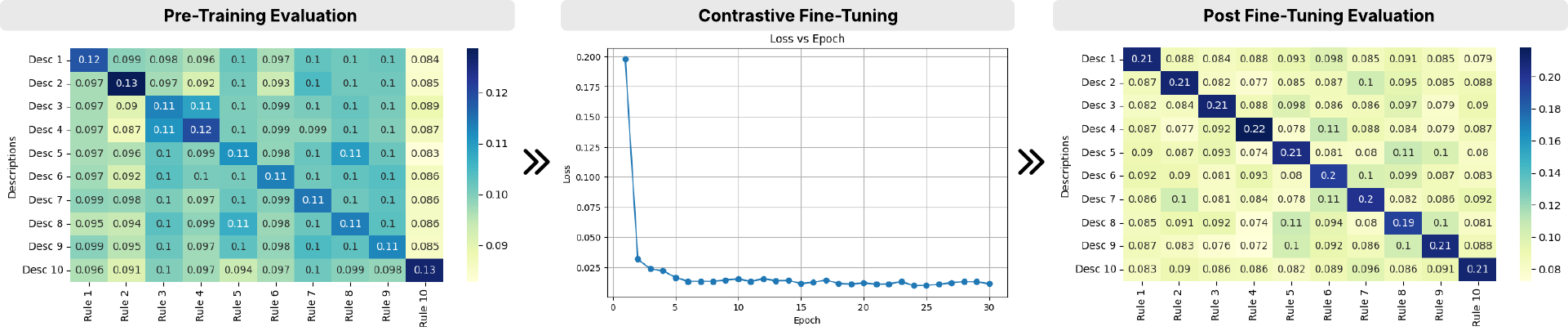}
        \caption{The correlation diagram shows the \textit{CTI–Rule Semantic Scorer}'s reliability in semantically mapping each CTI to its corresponding rule, with highest similarity scores (Sigmoid $\rightarrow$ Softmax) on the principal diagonal across 10 validation samples.}
        \label{fig:semantic_scorer_training}
    \end{figure*}

        \textbf{Contrastive fine-tuning (CFT):} For a batch of $N$ pairs we minimize the InfoNCE loss so the highest cosine falls on the diagonal pair~\cite{wu2021rethinking}:
        \begin{equation}\label{eq:infonce}
            \mathcal{L}_{\text{contrastive}} = -\sum_{i=1}^{N} \log \frac{\exp\!\big(\cos(\mathbf{z}_{\cti_i}, \mathbf{z}_{\idsrule_i}) / \tau\big)}{\sum_{j=1}^{N} \exp\!\big(\cos(\mathbf{z}_{\cti_i}, \mathbf{z}_{\idsrule_j}) / \tau\big)},
        \end{equation}
        with temperature $\tau$ and full objective $\mathcal{L}=\mathcal{L}_{\text{contrastive}}+\beta\,\mathcal{L}_{\text{KL}}$ ($\beta$-VAE~\cite{burgess2018understanding}). Minimizing it drives $\Phi(\cti_i,\idsrule_i)\to1$ for aligned pairs and $\to0$ otherwise [Eq.~\eqref{eq:phi}], the $\epsilon\to1$ desideratum of Eq.~\eqref{eq:scorer}. Both bi- and dual-encoder configurations train well under CFT [Fig.~\ref{fig:semantic_scorer_training}], with the bi-encoder (all-mpnet-base-v2\footnote{huggingface.co/sentence-transformers/all-mpnet-base-v2}) performing slightly better; we adopt it as the default $\Phi$, a single model serving both retrieval [Eq.~\eqref{eq:retrieval}] and validation [Eq.~\eqref{eq:accept}] [see Appendices~\ref{app:latent_consistency}--\ref{app:latent_consistency_validation} for further analysis].

    \subsection{\falcon Implementation and Module Interaction}\label{subsection:implementation}
        To show \falcon in practice, we trace a YARA ($\idsrule_i$) generation use case from a CTI input ($\cti_i$) through validation.

        \subsubsection{CTI Ingestion and Rule Generation}
            The process begins with a CTI input of signatures and behavior descriptors, e.g., a malware sample with known headers and behavior. The \textit{Relevant IDS Rule Retriever} surfaces semantically similar rules ($\idsrule^e_i$) via the semantic scorer and a preset threshold. A one-shot generation prompt tailored to the target format (e.g., Snort, YARA) combines $\instruction$, $\cti_i$, and $\idsrule^e_i$, and the \textit{Rule Generator LLM Agent} emits a candidate rule ($\idsrule_i$). See Appendix [Box~\ref{box:gen_instruction}] for full $\instruction$ and end-to-end use-case example.

            \begin{tcolorbox}[enhanced,attach boxed title to top center={yshift=-3mm,yshifttext=-1mm},
                colback=white,colframe=gray!75!black,colbacktitle=gray!80!black, title=Cyber Threat Intelligence $(\cti_i)$,
                  boxed title style={size=small,colframe=gray!90!black}, left=0.5mm, right=0.5mm, boxrule=0.75pt ]
                  \small
                \textbf{Threat Name:} \textcolor{teal}{HackTool\_MSIL\_CoreHound}\\
                \textbf{Threat Category:}\\
                \textcolor{teal}{
                    -- Malware / HackTool\\
                    -- .NET-based Threat ...\\
                    }
                \textbf{Indicators of Compromise (IoCs):}\\
                \textcolor{teal}{
                    -- TypeLibGUID / ProjectGuid: 1fff2aee-a540-4613-94ee-...\\
                    -- MD5 Hash: dd8805d0e470e59b829d98397507d8c2 ...\\
                    }
                \textbf{Observed Behavior:}\\
                \textcolor{teal}{
                    1. Windows PE file by MZ (0x5A4D) header at file beginning.\\
                    2. PE signature (0x00004550) at specified localtion in header ...
                    }
            \end{tcolorbox}

            \subsubsection{Generated IDS Rule}
                The agent's output ($\idsrule_i$) encodes the threat-prevention logic derived from $\cti_i$ and $\idsrule_i^e$, typically including protocol, domain-matching, and content-matching elements. As an unvalidated candidate, it must pass syntactic, semantic (functional), and performance checks plus SOC review before deployment.

            \subsubsection{Syntactic Validation}
                $\idsrule_i$ is first sent to the Syntax Validator, which parses it for structural correctness. Errors return Syntactic Validator Feedback ($\feedback^s$) for regeneration; otherwise the rule advances to semantic validation [Box~\ref{box:syntactic_fb}].

            \subsubsection{Semantic Validation}
                A valid rule proceeds to the Semantic Validator, where the \textit{CTI-Rule Semantic Scorer} [see Section~\ref{subsection:semantic_scorer}] quantifies its functional similarity with $\cti_i$. The \textit{Semantic Analysis LLM Agent}, driven by this score, flags residual inconsistencies, such as missing indicators, incorrect protocols, or irrelevant payloads, and, if critical, formulates targeted feedback ($\feedback^f$) for the \textit{Rule Generator}, prompting a revised generation cycle [Box~\ref{box:semantic_fb}].

            \subsubsection{Performance Validation}
                The \textit{Performance Validator} then checks runtime efficiency, including regex use, rule complexity, match execution time, and mapping to existing rules ($\idsrule^e_i$). Inefficient rules are flagged and routed back with performance-specific feedback ($\feedback^p$) [Box~\ref{box:performance_fb}].

            \subsubsection{Cybersecurity Analyst Feedback}
                Validated rules and the consolidated Analyst Note ($\feedback^f \cup \feedback^p$) are forwarded to a \textit{Cybersecurity Analyst} for manual review and approval before deployment. The analyst may also request regeneration under new constraints via feedback ($\feedback^a$), enforcing organizational policy and compliance.

                \begin{tcolorbox}[enhanced,attach boxed title to top center={yshift=-3mm,yshifttext=-1mm},
                    colback=white,colframe=gray!75!black,colbacktitle=gray!80!black, title=Final YARA Rule $(\idsrule_i)$,
                      boxed title style={size=small,colframe=gray!90!black}, left=0.5mm, right=0.5mm, boxrule=0.75pt ]
                      \small
                    \textbf{rule HackTool\_MSIL\_CoreHound \{ }\\
                    \textbf{meta:}\\
                    \textcolor{teal}{
                        \textcolor{white}{----}description = "The TypeLibGUID present in a .NET binary ..."\\
                        \textcolor{white}{----}md5 = "dd8805d0e470e59b829d98397507d8c2"\\
                    }
                    \textbf{strings:}\\
                    \textcolor{teal}{
                        \textcolor{white}{----}typelibguid0 = ``1fff2aee-a540-..." ascii nocase wide\\
                    }
                    \textbf{condition:}\\
                    \textcolor{teal}{
                        \textcolor{white}{----}(uint16(0) == 0x5A4D and uint32(...) and any of them\\
                        \textcolor{black}{\textbf{\}}}
                    }
                \end{tcolorbox}

                This modular, agent-driven design lets each component specialize in a distinct function and enables iterative refinement. Combining LLM-driven generation, parser-based syntax checks, functional-consistency scoring, and performance profiling makes \falcon adaptable yet efficient at producing high-quality IDS rules at scale.
    
\section{Experiment \& Evaluation}\label{section:experiment}
    
    In this section, we present the experiments conducted to validate our proposed \falcon framework. We designed three types of evaluation: training and assessing the performance of the \textit{CTI-Rule Semantic Scorer} [Sec. \ref{subsection:semantic_scorer_evaluation}], evaluation of \textit{Rule Generator LLM Agent} [Sec. \ref{subsection:rule_generator_evaluation}], and end-to-end qualitative validation of \falcon pipeline [Sec. \ref{subsection:pipeline_qualitative}]. 
    

    \subsection{Data Description and Experiment Setup}
        To train and evaluate our \textit{CTI-Rule Semantic Scorer} model and end-to-end \falcon pipeline, we collected 4017 Snort\footnote{Snort (Community): \url{snort.org}} and 4587 YARA\footnote{YARA: \url{github.com/Yara-Rules/rules}} rules ($\idsrule_i$) from open-source repositories and threat intelligence datasets. Two CTI instances ($\cti_i$) were carefully generated for each rule, reflecting distinct but relevant threat behavior and signature descriptions. Additionally, we generated a list of relevant but outdated rules ($\idsrule_i^e$) for each rule to test out \textit{CTI-Rule Semantic Scorer} as a retriever. This resulted in 8034 Snort and 9174 YARA CTI-Rule pairs and 15217 snort and 25875 YARA deployed outdated rules ($\idsrule^e$) for the retriever assessment. The CTI-Rule dataset was then split into 90\% for \textit{CTI-Rule} model training and 10\% for testing (802 Snort and 916 YARA), ensuring balanced representation across both types of IDS rules. Apart from testing set, 60 Snort and 60 YARA were separately designed as validation set for the overall pipeline's qualitative evaluation. Each rule in the validation set was categorized into one of three difficulty levels (\textit{Easy}, \textit{Medium}, or \textit{Hard}) based on complexity and length assessments performed by Subject Matter Experts (SMEs) which are authors in our case. These CTI and obsolete yet relevant IDS rules were designed to simulate real-world use cases where \falcon must generate and validate rules from novel CTI inputs and existing deployed rules. To simulate diverse operational contexts, CTI was initially tested in both semi-structured natural language and structured STIX 2.0~\cite{sadique2018automated} formats. Our preliminary observations indicated that semi-structured CTI led to more accurate generation of IDS rules (example in Section~\ref{section:aidrag_framework}), and we standardized subsequent evaluations on similar predefined CTI format. However, the CTI schema remains flexible and can be adapted to existing cybersecurity requirements, such as STIX or others. The \textit{CTI-Rule Semantic Scorer} models were built on top of pre-trained embedding models and trained using contrastive learning (details in Section \ref{subsection:semantic_scorer}) with different hyperparameter configurations, and early stopping based on validation loss. This selection is intentional, as our findings with LLM embeddings were poor and inefficient. All training was performed on six NVIDIA H200 GPUs, and dataset, code, preliminary, and complete experimental results are reported in the repository\footref{github}.

    \subsection{Semantic Scorer Model Evaluation}\label{subsection:semantic_scorer_evaluation}
        To validate our answer to \textbf{RQ-2} [Section~\ref{subsection:semantic_scorer}], we benchmark the \textit{CTI-Rule Semantic Scorer} against four baselines across its two roles, retrieval and ground-truth-free validation [Table~\ref{tab:scorer_full}]. The \textbf{Consistency} columns establish its validity as a validator: lacking a reference rule at test time, we check that scoring a generated rule against its CTI (CTI$\to$G) tracks scoring it against the ground-truth rule (GT$\to$G), via their mean absolute gap ($|\Delta|\!\downarrow$) and correlation ($r\!\uparrow$); per-platform scatter plots of this CTI$\to$G vs GT$\to$G agreement appear in Appendix~\ref{app:latent_consistency_validation}. The \textbf{Retriever} columns assess retrieval of relevant deployed rules by Recall@10 (R@10) and Mean Average Precision (MAP). The \textbf{Semantic} columns measure a generated rule's alignment with its CTI by diagonal recall (S.Rec), i.e., whether the principal-diagonal score is highest in its row, and a thresholded F1 (Thr) at the optimal post-scaling threshold. Per column, best shaded green and second-best blue.

        \begin{table*}[t]
          \centering
          \caption{Validation-set evaluation of semantic scorers. \textit{Consistency}: agreement of $\mathrm{score}(\text{CTI},\text{Gen})$ vs $\mathrm{score}(\text{GT},\text{Gen})$ over 802 Snort / 916 YARA CTIs $\times$ 10 LLMs ($|\Delta|\downarrow$, $r\uparrow$). \textit{Retriever}/\textit{Semantic}: ranking each CTI against the gold-rule pool (770 Snort / 871 YARA). Per column (within a family), best is \colorbox{softgreen}{green}, second \colorbox{softblue}{blue}.}\label{tab:scorer_full}
          \scriptsize
          \renewcommand{\arraystretch}{1.2}
          \setlength{\tabcolsep}{5pt}
          \begin{tabular}{ll|ll|ll|ll|ll}
          \hline
          \rowcolor{lightgray}
          \multicolumn{2}{l|}{} & \multicolumn{4}{c|}{Consistency [0-1]} & \multicolumn{2}{c|}{Retriever (\%)} & \multicolumn{2}{c}{Semantic [0-1]} \\ \hline
          \multicolumn{1}{l|}{Scorer} & Case & \multicolumn{1}{c}{CTI$\to$G} & \multicolumn{1}{c}{GT$\to$G} & $|\Delta|\downarrow$ & \multicolumn{1}{c|}{$r\uparrow$} & R@10 & \multicolumn{1}{c|}{MAP} & S.Rec & Thr \\ \hline\hline
          \multicolumn{1}{l|}{\multirow{2}{*}{CTI-Rule (Ours)}} & Snort & 0.952 & 0.966 & \cellcolor{softgreen}0.023 & \multicolumn{1}{l|}{\cellcolor{softgreen}0.888} & \cellcolor{softgreen}100.0 & \multicolumn{1}{l|}{\cellcolor{softgreen}99.8} & \cellcolor{softgreen}0.995 & \cellcolor{softgreen}0.910 \\ \cline{2-10}
          \multicolumn{1}{l|}{} & YARA & 0.964 & 0.971 & \cellcolor{softgreen}0.012 & \multicolumn{1}{l|}{\cellcolor{softgreen}0.974} & \cellcolor{softgreen}100.0 & \multicolumn{1}{l|}{\cellcolor{softgreen}100.0} & \cellcolor{softgreen}1.000 & \cellcolor{softgreen}0.935 \\ \hline
          \multicolumn{1}{l|}{\multirow{2}{*}{BERT-F1}} & Snort & 0.428 & 0.783 & 0.355 & \multicolumn{1}{l|}{0.428} & 72.0 & \multicolumn{1}{l|}{62.1} & 0.565 & 0.545 \\ \cline{2-10}
          \multicolumn{1}{l|}{} & YARA & 0.578 & 0.724 & 0.163 & \multicolumn{1}{l|}{0.447} & 93.0 & \multicolumn{1}{l|}{90.4} & 0.888 & 0.765 \\ \hline
          \multicolumn{1}{l|}{\multirow{2}{*}{RAGAS}} & Snort & 0.675 & 0.944 & 0.269 & \multicolumn{1}{l|}{0.511} & \cellcolor{softblue}99.9 & \multicolumn{1}{l|}{97.0} & 0.949 & \cellcolor{softblue}0.850 \\ \cline{2-10}
          \multicolumn{1}{l|}{} & YARA & 0.734 & 0.885 & 0.152 & \multicolumn{1}{l|}{0.410} & \cellcolor{softblue}99.0 & \multicolumn{1}{l|}{89.8} & 0.845 & \cellcolor{softblue}0.800 \\ \hline
          \multicolumn{1}{l|}{\multirow{2}{*}{BM25}} & Snort & 0.305 & 0.107 & 0.198 & \multicolumn{1}{l|}{\cellcolor{softblue}0.743} & 99.5 & \multicolumn{1}{l|}{96.0} & 0.940 & 0.280 \\ \cline{2-10}
          \multicolumn{1}{l|}{} & YARA & 0.126 & 0.099 & \cellcolor{softblue}0.030 & \multicolumn{1}{l|}{\cellcolor{softblue}0.946} & 92.0 & \multicolumn{1}{l|}{88.8} & 0.867 & 0.135 \\ \hline
          \multicolumn{1}{l|}{\multirow{2}{*}{TF-IDF + Cosine}} & Snort & 0.566 & 0.622 & \cellcolor{softblue}0.123 & \multicolumn{1}{l|}{0.614} & 99.8 & \multicolumn{1}{l|}{\cellcolor{softblue}97.0} & \cellcolor{softblue}0.954 & 0.500 \\ \cline{2-10}
          \multicolumn{1}{l|}{} & YARA & 0.590 & 0.776 & 0.205 & \multicolumn{1}{l|}{0.683} & 98.0 & \multicolumn{1}{l|}{\cellcolor{softblue}97.5} & \cellcolor{softblue}0.971 & 0.725 \\ \hline
          \end{tabular}
        \end{table*}

    \subsection{Rule Generator LLM Agent Evaluation}\label{subsection:rule_generator_evaluation}
        \textbf{RQ-1 and RQ-3.} Table~\ref{tab:full_all} jointly evaluates rule-generation feasibility (RQ-1) and self-reflective refinement (RQ-3) over 18 LLMs, 14 general-purpose models across Small (S, 3--4B), Medium (M, 8--30B), Large (L, 32--49B), and Extra-Large (XL, 117--128B) scales plus four cyber-security fine-tuned models (Cyb), for both Snort (NIDS) and YARA (HIDS). Each model is run in two modes: \textit{Single}, a one-shot generation with no validator feedback (the RQ-1 baseline), and \textit{Chain}, the full validator-chained self-reflective loop of Section~\ref{section:aidrag_framework} in which syntactic, semantic, and performance feedback drives iterative regeneration (RQ-3). Generated rules are scored by our CTI-Rule scorer (sigmoid functional alignment), RAGAS~\cite{es2024ragas} (OpenAI-embedding cosine), and BERT-F1~\cite{zhang2019bertscore} (baseline-rescaled BERTScore), each against the input CTI (CTI vs Gen) and the ground-truth rule (Gen vs GT); \textbf{Pass} counts generations clearing all three validators (out of 802 Snort / 916 YARA). To test whether agentic reasoning adds value \emph{on top of} self-reflection, the five reasoning-capable models (Qwen3-4B/14B/32B, Nemotron-Super-49B, and GPT-OSS-117B) are additionally run with thinking enabled and disabled (Th on/off).

        For \textbf{RQ-1}, even in one-shot \textit{Single} mode, models across all scales transform heterogeneous CTI into relevant, deployable rules, with the CTI-Rule score holding a stable $\approx\!0.72$ where lexical BERT-F1 swings widely. For \textbf{RQ-3}, \textit{Chain} consistently raises the Pass rate over \textit{Single} (e.g., Qwen3-32B Snort $451\!\to\!649$, Mistral-Medium YARA $597\!\to\!912$), confirming that structured validator feedback yields measurable gains over one-shot generation. Enabling thinking on the five reasoning-capable models produces only marginal, inconsistent changes on top of \textit{Chain}, indicating that the validator-driven self-reflection, rather than reasoning alone, is the primary driver of improvement.

        \begin{table*}[t]
        \centering
        \setlength{\tabcolsep}{3pt}
        \def\arraystretch{1.12}
        \caption{Comprehensive evaluation of all LLMs across generation modes (Single / Chain) and reasoning states (thinking on/off), for both rule families. CTI-Rule = our CTI-Rule semantic scorer; RAGAS = OpenAI-embedding; Bert-F1 = baseline-rescaled BERTScore. Per column, best is \colorbox{softgreen}{green} and second-best \colorbox{softblue}{blue}. \textcolor{red}{Red} labels (\texttt{Cyb}) are cyber-security fine-tuned models.}
        \label{tab:full_all}
        \tiny
        \resizebox{\textwidth}{!}{%
        \begin{tabular}{l|l|r|c|c|ccc|ccc|c|ccc|ccc|c}
        \hline\rowcolor{lightgray}
        \multicolumn{5}{c|}{} & \multicolumn{3}{c|}{\textbf{CTI vs Gen. Snort}} & \multicolumn{3}{c|}{\textbf{Gen. Snort vs GT Snort}} & \multicolumn{1}{c|}{\textbf{Pass}} & \multicolumn{3}{c|}{\textbf{CTI vs Gen. YARA}} & \multicolumn{3}{c|}{\textbf{Gen. YARA vs GT YARA}} & \multicolumn{1}{c}{\textbf{Pass}}\\
        \rowcolor{lightgray}
        Sz & Model & Param & Mode & Th & CTI-Rule & Ragas & Bert-F1 & CTI-Rule & Ragas & Bert-F1 & Snort & CTI-Rule & Ragas & Bert-F1 & CTI-Rule & Ragas & Bert-F1 & YARA\\\hline\hline
        \multirow{2}{*}{S} & \multirow{2}{*}{Granite-4.1} & \multirow{2}{*}{3B} & Single & -- & 0.720 & 0.700 & \cellcolor{softgreen}-0.146 & 0.723 & 0.952 & 0.458 & 682/802 & 0.720 & 0.690 & 0.016 & 0.722 & 0.884 & 0.258 & 289/916\\
          &  &  & Chain & -- & 0.720 & 0.700 & \cellcolor{softblue}-0.147 & 0.723 & 0.952 & 0.457 & 692/802 & 0.717 & 0.688 & 0.023 & 0.720 & 0.881 & 0.268 & 427/916\\
        \cline{1-19}
        \multirow{2}{*}{S} & \multirow{2}{*}{Llama-3.2} & \multirow{2}{*}{3B} & Single & -- & 0.718 & 0.728 & -0.179 & 0.721 & 0.935 & 0.452 & 688/802 & 0.721 & 0.714 & 0.127 & 0.722 & 0.886 & 0.392 & \textbf{162}/916\\
          &  &  & Chain & -- & 0.719 & 0.730 & -0.179 & 0.721 & 0.937 & 0.452 & 744/802 & 0.721 & 0.715 & 0.116 & 0.722 & 0.886 & 0.379 & \textbf{477/916}\\
        \cline{1-19}
        \multirow{2}{*}{S} & \multirow{2}{*}{Phi-4-mini} & \multirow{2}{*}{4B} & Single & -- & 0.717 & 0.720 & -0.206 & 0.719 & 0.924 & 0.408 & 555/802 & 0.720 & 0.750 & 0.187 & 0.721 & 0.881 & 0.393 & 267/916\\
          &  &  & Chain & -- & 0.718 & 0.718 & -0.205 & 0.720 & 0.920 & 0.401 & 592/802 & 0.720 & 0.750 & 0.183 & 0.721 & 0.881 & 0.386 & 327/916\\
        \cline{1-19}
        \multirow{4}{*}{S} & \multirow{4}{*}{Qwen3} & \multirow{4}{*}{4B} & Single & off & 0.720 & 0.728 & -0.162 & 0.724 & 0.960 & 0.527 & 782/802 & 0.724 & 0.766 & 0.172 & 0.726 & 0.883 & 0.420 & 524/916\\
          &  &  & Single & on & 0.722 & 0.728 & -0.154 & 0.725 & 0.962 & 0.527 & 750/802 & 0.724 & 0.762 & 0.156 & 0.726 & 0.888 & 0.406 & 497/916\\
          &  &  & Chain & off & 0.720 & 0.728 & -0.162 & 0.724 & 0.960 & 0.528 & 782/802 & 0.724 & 0.765 & 0.173 & 0.726 & 0.885 & 0.421 & 715/916\\
          &  &  & Chain & on & 0.722 & 0.728 & -0.153 & 0.725 & 0.962 & 0.527 & 765/802 & 0.724 & 0.761 & 0.154 & 0.725 & 0.887 & 0.400 & 657/916\\
        \cline{1-19}
        \multirow{2}{*}{S} & \multirow{2}{*}{Gemma-3} & \multirow{2}{*}{4B} & Single & -- & 0.716 & 0.722 & -0.191 & 0.719 & 0.919 & 0.417 & 583/802 & 0.714 & 0.743 & 0.135 & 0.715 & 0.879 & 0.388 & 682/916\\
          &  &  & Chain & -- & 0.715 & 0.720 & -0.192 & 0.718 & 0.917 & 0.415 & 584/802 & 0.711 & 0.737 & 0.125 & 0.712 & 0.872 & 0.375 & 721/916\\
        \cline{1-19}
        \multirow{2}{*}{M} & \multirow{2}{*}{Granite-4.1} & \multirow{2}{*}{8B} & Single & -- & 0.722 & 0.724 & -0.162 & 0.725 & 0.960 & 0.534 & 639/802 & 0.725 & 0.740 & 0.153 & \cellcolor{softblue}0.727 & 0.901 & 0.456 & \textbf{437}/916\\
          &  &  & Chain & -- & 0.721 & 0.718 & -0.164 & 0.725 & 0.947 & 0.517 & 646/802 & 0.725 & 0.737 & 0.150 & 0.727 & 0.900 & 0.446 & \textbf{685/916}\\
        \cline{1-19}
        \multirow{2}{*}{M} & \multirow{2}{*}{Llama-3.1} & \multirow{2}{*}{8B} & Single & -- & \cellcolor{softblue}0.723 & 0.731 & -0.164 & 0.726 & 0.963 & 0.558 & 781/802 & 0.724 & 0.726 & 0.131 & 0.725 & 0.903 & 0.421 & \textbf{516}/916\\
          &  &  & Chain & -- & 0.723 & \cellcolor{softblue}0.732 & -0.161 & 0.726 & 0.964 & 0.561 & 790/802 & 0.724 & 0.727 & 0.129 & 0.725 & \cellcolor{softblue}0.904 & 0.423 & \textbf{842/916}\\
        \cline{1-19}
        \multirow{4}{*}{M} & \multirow{4}{*}{Qwen3} & \multirow{4}{*}{14B} & Single & off & \cellcolor{softgreen}0.724 & 0.729 & -0.164 & \cellcolor{softblue}0.727 & 0.966 & 0.572 & 761/802 & \cellcolor{softblue}0.726 & 0.773 & 0.201 & 0.727 & 0.891 & 0.458 & 692/916\\
          &  &  & Single & on & 0.722 & 0.731 & -0.163 & 0.726 & 0.969 & 0.569 & 789/802 & 0.725 & 0.773 & 0.191 & 0.726 & 0.889 & 0.443 & 549/916\\
          &  &  & Chain & off & 0.724 & 0.730 & -0.163 & 0.727 & 0.967 & 0.573 & 763/802 & 0.726 & 0.773 & 0.199 & 0.727 & 0.890 & 0.456 & 804/916\\
          &  &  & Chain & on & 0.722 & 0.731 & -0.163 & 0.726 & 0.969 & 0.569 & 795/802 & 0.725 & 0.771 & 0.188 & 0.726 & 0.889 & 0.436 & 798/916\\
        \cline{1-19}
        \multirow{2}{*}{M} & \multirow{2}{*}{Mistral-Small} & \multirow{2}{*}{24B} & Single & -- & 0.723 & 0.717 & -0.164 & 0.727 & 0.964 & 0.572 & 785/802 & 0.724 & 0.749 & 0.147 & 0.726 & 0.894 & 0.423 & 537/916\\
          &  &  & Chain & -- & 0.723 & 0.718 & -0.163 & 0.727 & 0.965 & 0.574 & \cellcolor{softblue}\textbf{799/802} & 0.723 & 0.749 & 0.144 & 0.725 & 0.890 & 0.415 & \textbf{864/916}\\
        \cline{1-19}
        \multirow{2}{*}{M} & \multirow{2}{*}{Granite-4.1} & \multirow{2}{*}{30B} & Single & -- & 0.723 & 0.722 & -0.155 & 0.726 & 0.966 & 0.544 & 790/802 & 0.723 & 0.702 & 0.145 & 0.726 & \cellcolor{softgreen}0.905 & 0.474 & 490/916\\
          &  &  & Chain & -- & 0.723 & 0.722 & -0.155 & 0.726 & 0.966 & 0.545 & 794/802 & 0.723 & 0.701 & 0.139 & 0.726 & 0.903 & 0.468 & 597/916\\
        \cline{1-19}
        \multirow{4}{*}{L} & \multirow{4}{*}{Qwen3} & \multirow{4}{*}{32B} & Single & off & 0.723 & 0.727 & -0.169 & 0.727 & 0.961 & 0.583 & 451/802 & 0.726 & 0.767 & \cellcolor{softgreen}0.206 & \cellcolor{softgreen}0.728 & 0.898 & \cellcolor{softgreen}0.495 & 648/916\\
          &  &  & Single & on & 0.722 & 0.729 & -0.166 & 0.725 & 0.969 & 0.574 & 756/802 & 0.726 & \cellcolor{softgreen}0.776 & 0.193 & 0.727 & 0.891 & 0.432 & 604/916\\
          &  &  & Chain & off & 0.723 & 0.726 & -0.169 & 0.727 & 0.962 & 0.580 & 649/802 & 0.726 & 0.765 & \cellcolor{softblue}0.205 & 0.728 & 0.898 & \cellcolor{softblue}0.488 & 879/916\\
          &  &  & Chain & on & 0.722 & 0.729 & -0.167 & 0.726 & 0.969 & 0.574 & 774/802 & 0.726 & \cellcolor{softblue}0.775 & 0.192 & 0.727 & 0.892 & 0.428 & 889/916\\
        \cline{1-19}
        \multirow{4}{*}{L} & \multirow{4}{*}{Nemotron-Super} & \multirow{4}{*}{49B} & Single & off & 0.722 & 0.730 & -0.170 & 0.726 & 0.967 & 0.567 & 752/802 & 0.725 & 0.760 & 0.158 & 0.727 & 0.894 & 0.439 & 433/916\\
          &  &  & Single & on & 0.718 & 0.726 & -0.177 & 0.721 & 0.963 & 0.520 & 718/802 & 0.724 & 0.749 & 0.137 & 0.725 & 0.888 & 0.396 & 383/916\\
          &  &  & Chain & off & 0.722 & 0.731 & -0.170 & 0.726 & 0.967 & 0.565 & 758/802 & 0.725 & 0.758 & 0.152 & 0.726 & 0.893 & 0.427 & \textbf{795/916}\\
          &  &  & Chain & on & 0.718 & 0.726 & -0.178 & 0.722 & 0.963 & 0.519 & 740/802 & 0.724 & 0.753 & 0.135 & 0.726 & 0.893 & 0.398 & \textbf{733/916}\\
        \cline{1-19}
        \multirow{4}{*}{XL} & \multirow{4}{*}{GPT-OSS} & \multirow{4}{*}{117B} & Single & off & 0.722 & 0.731 & -0.181 & 0.726 & 0.966 & 0.582 & 611/802 & 0.726 & 0.768 & 0.183 & 0.727 & 0.883 & 0.435 & 569/916\\
          &  &  & Single & on & 0.722 & 0.730 & -0.176 & 0.726 & \cellcolor{softblue}0.971 & \cellcolor{softblue}0.595 & 699/802 & 0.726 & 0.771 & 0.193 & 0.728 & 0.887 & 0.458 & 496/916\\
          &  &  & Chain & off & 0.722 & 0.732 & -0.180 & 0.726 & 0.970 & 0.582 & 748/802 & \cellcolor{softgreen}0.727 & 0.769 & 0.187 & 0.728 & 0.887 & 0.440 & \cellcolor{softblue}\textbf{894/916}\\
          &  &  & Chain & on & 0.723 & 0.731 & -0.177 & 0.726 & \cellcolor{softgreen}0.972 & 0.591 & 798/802 & 0.725 & 0.770 & 0.194 & 0.727 & 0.885 & 0.455 & 736/916\\
        \cline{1-19}
        \multirow{2}{*}{XL} & \multirow{2}{*}{Mistral-Medium} & \multirow{2}{*}{128B} & Single & -- & 0.724 & 0.731 & -0.159 & \cellcolor{softgreen}0.728 & 0.972 & \cellcolor{softgreen}0.601 & \cellcolor{softgreen}\textbf{800/802} & 0.726 & 0.750 & 0.156 & 0.728 & 0.899 & 0.453 & 597/916\\
          &  &  & Chain & -- & 0.724 & 0.731 & -0.159 & 0.728 & 0.971 & 0.601 & \textbf{800/802} & 0.726 & 0.749 & 0.149 & 0.727 & 0.898 & 0.439 & \cellcolor{softgreen}\textbf{912/916}\\
        \cline{1-19}
        \multirow{2}{*}{\textcolor{red}{Cyb}} & \multirow{2}{*}{\textcolor{red}{Lily-Cyber}} & \multirow{2}{*}{7B} & Single & -- & 0.714 & 0.700 & -0.230 & 0.717 & 0.866 & 0.292 & 756/802 & 0.709 & 0.690 & -0.045 & 0.710 & 0.832 & 0.169 & 317/916\\
          &  &  & Chain & -- & 0.714 & 0.700 & -0.231 & 0.717 & 0.865 & 0.292 & 757/802 & 0.699 & 0.679 & 0.011 & 0.700 & 0.819 & 0.232 & 390/916\\
        \cline{1-19}
        \multirow{2}{*}{\textcolor{red}{Cyb}} & \multirow{2}{*}{\textcolor{red}{SecGPT}} & \multirow{2}{*}{7B} & Single & -- & 0.722 & \cellcolor{softgreen}0.736 & -0.170 & 0.725 & 0.959 & 0.532 & 764/802 & 0.719 & 0.758 & 0.147 & 0.720 & 0.871 & 0.397 & 570/916\\
          &  &  & Chain & -- & 0.722 & 0.736 & -0.171 & 0.725 & 0.959 & 0.530 & 775/802 & 0.719 & 0.758 & 0.146 & 0.720 & 0.871 & 0.397 & 748/916\\
        \cline{1-19}
        \multirow{2}{*}{\textcolor{red}{Cyb}} & \multirow{2}{*}{\textcolor{red}{Foundation-Sec}} & \multirow{2}{*}{8B} & Single & -- & 0.723 & 0.729 & -0.166 & 0.726 & 0.967 & 0.569 & \textbf{791}/802 & 0.725 & 0.763 & 0.177 & 0.726 & 0.893 & 0.450 & 766/916\\
          &  &  & Chain & -- & 0.723 & 0.729 & -0.165 & 0.726 & 0.967 & 0.571 & \textbf{797/802} & 0.724 & 0.763 & 0.174 & 0.726 & 0.893 & 0.447 & 829/916\\
        \cline{1-19}
        \multirow{2}{*}{\textcolor{red}{Cyb}} & \multirow{2}{*}{\textcolor{red}{Llama-Primus}} & \multirow{2}{*}{8B} & Single & -- & 0.722 & 0.713 & -0.191 & 0.726 & 0.929 & 0.506 & 751/802 & 0.725 & 0.735 & 0.137 & 0.726 & 0.893 & 0.412 & \textbf{783}/916\\
          &  &  & Chain & -- & 0.722 & 0.713 & -0.191 & 0.726 & 0.929 & 0.504 & 754/802 & 0.725 & 0.735 & 0.136 & 0.726 & 0.893 & 0.411 & \textbf{830/916}\\
        \cline{1-19}
        \hline
        \end{tabular}}
        \end{table*}

        \begin{table}[t]
          \centering
          \caption{Paired mean improvement, averaged over all matched model$\times$family$\times$perspective configs. $\Delta$: absolute change; \%: relative change.}
          \label{tab:paired_improvement}
          \scriptsize
          \renewcommand{\arraystretch}{1.2}
          \setlength{\tabcolsep}{4pt}
          \begin{tabular}{l|ll|ll}
          \hline
          \rowcolor{lightgray}
          & \multicolumn{2}{c|}{All-3 pass} & \multicolumn{2}{c}{Syntax pass} \\ \hline
          Comparison & \multicolumn{1}{c}{$\Delta$} & \multicolumn{1}{c|}{\%} & \multicolumn{1}{c}{$\Delta$} & \multicolumn{1}{c}{\%} \\ \hline\hline
          Chain-validator vs single-shot & $+0.130$ & $+18.0\%$ & $+0.127$ & $+17.2\%$ \\ \hline
          Thinking on vs off & $-0.001$ & $-0.1\%$ & $+0.006$ & $+0.7\%$ \\ \hline
          \end{tabular}
        \end{table}

        \begin{figure*}[!ht]
            \centering \includegraphics[width=0.9\textwidth]{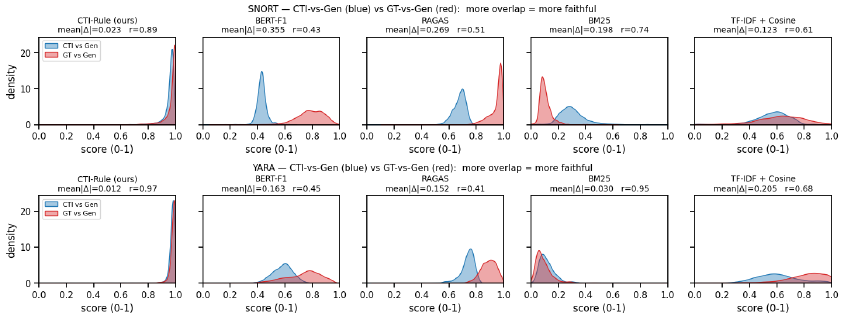}
            \caption{The diagram illustrates the reliability of our \falcon CTI-Rule scorer model. This is evidenced by the minimum variance ($|\Delta|$) and maximum correlation ($r$) between scores derived from CTI inputs, generated rules, and ground truth rules.}
            \label{fig:semantic_scorer_final_eval}
        \end{figure*}
    
        \begin{figure*}[!ht]
            \centering \includegraphics[width=0.9\textwidth]{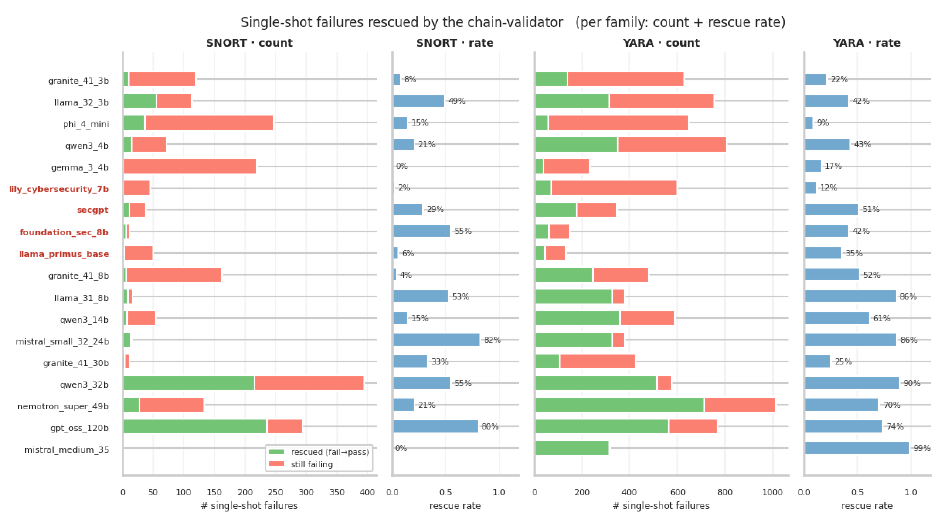}
            \caption{Single-shot failures rescued by the chain-validator, per model (Snort, YARA). \emph{Count} panels: rescued (green, fail$\to$pass) vs.\ still failing (red); \emph{rate} panels: rescue rate (blue). Red labels denote cyber-tuned models.}
            \label{fig:e2e_semantic_scorer_final_eval}
        \end{figure*}

        \subsection{End-to-End \falcon Pipeline Evaluation}\label{subsection:pipeline_qualitative}

            To address \textbf{RQ-3}, i.e., whether an LLM can interpret validator feedback to improve a generated rule, we close the loop between the validation and generation phases and re-run the same 10\% evaluation split (802 Snort, 916 YARA) with the best-performing LLM in each size category [Table~\ref{tab:full_all}], plotting the refined semantic score (orange) against one-shot generation (blue) for the runs that required refinement [Fig.~\ref{fig:e2e_semantic_scorer_final_eval}]. Aggregated across all matched model$\times$family$\times$perspective configurations, chain-validator refinement raises the semantic score by $+0.130$ ($+18.0\%$) over single-shot, whereas enabling agentic reasoning shifts it by a negligible $\approx\!0\%$ ($-0.1\%$ all-validator, $+0.7\%$ syntax-only) [Table~\ref{tab:paired_improvement}], confirming that the validator loop, not reasoning, drives the gain. This lift is measured from the initial one-shot rule, including sub-threshold cases, to the refined output; accepted rules themselves cluster near the validator thresholds. A per-validator breakdown of the single-shot failures rescued by the chain, together with the thinking on/off comparison, appears in Appendix~\ref{app:validator_ablation}.
        
            To corroborate the quantitative results, we qualitatively assess end-to-end feasibility on 60 isolated validation samples. Three cyber-security Subject Matter Experts (SMEs) independently rate 60 generated rule on a Likert scale: non-match (0), syntactically correct (1), semantically correct (2), and performance-optimized (3) [Table~\ref{tab:qualitative}]. Scores are aggregated per difficulty category and normalized to $[0\text{--}1]$ to measure consensus. We observe \textbf{84\%} inter-rater agreement, indicating the consistency and reliability of \falcon.
        
            \begin{table}[h]
                \centering
                \def\arraystretch{1.2}
                \caption{Qualitative evaluation of \falcon rule generation. Scores reflect inter-rater agreement $[0\text{--}1]$ among SMEs.}
                \label{tab:qualitative}
                \scriptsize
                \begin{tabular}{l|l|l|l|l}
                \hline
                \rowcolor{lightgray}
                                Medium-Platform    & Difficulty & Llama     & Mistral     & Granite       \\ \hline
                \hline
                \multirow{3}{*}{NIDS-Snort} & Easy     & 1.00      & 1.00        & 1.00          \\ \cline{2-5}
                                            & Medium   & 0.98      & 0.98        & 1.00          \\ \cline{2-5}
                                            & Hard     & 0.95      & 0.93        & 1.00          \\ \hline
                \hline
                \multirow{3}{*}{HIDS-YARA}  & Easy     & 0.95      & 0.98        & 1.00          \\ \cline{2-5}
                                            & Medium   & 0.86      & 0.86        & 0.92          \\ \cline{2-5}
                                            & Hard     & 0.95      & 0.93        & 0.96          \\ \hline
                \end{tabular}
            \end{table}

\subsection{Discussion}

    From experimental results, we draw following observations:

    \textbf{(1) Retrieval is competitive but validation is not.} On the retrieval task, the CTI-Rule model attains near-perfect Recall@10 and MAP [Table~\ref{tab:scorer_full}], but sparse and dense baselines (TF-IDF, BM25, RAGAS) remain close because deployed rules and their CTIs share substantial token overlap. Retrieval is therefore best served by an efficient ensemble (sparse + dense) re-ranker. The picture inverts for ground-truth-free validation: lexical and embedding baselines collapse and the CTI-Rule scorer becomes indispensable, as seen in the Consistency and thresholded-F1 columns of Table~\ref{tab:scorer_full}.

    \textbf{(2) Larger models has more coverage (RQ-1).} Across the full suite [Table~\ref{tab:full_all}], models from 3B to 128B all produce syntactically, semantically, and operationally valid rules, but the number of CTIs successfully translated (\textbf{Pass}) grows with model scale: the largest model, Mistral-Medium (128B), clears the most ($800/802$ Snort, $912/916$ YARA), whereas the small 3--4B models (e.g., Phi-4-mini, Llama-3.2) translate markedly fewer, especially on YARA. Because every accepted rule must clear the same validator thresholds, post-pass similarity scores cluster in a narrow band; the discriminating quantity is therefore \emph{coverage} (Pass), not score magnitude.

    \textbf{(3) CTI-Rule scorer similarity is logical (RQ-2).} Comparing semantic-similarity metrics, RAGAS tends to overestimate and BERT-F1 to underestimate alignment, and both fluctuate sharply across models and rule families. Our CTI-Rule scorer instead stays stable ($\approx\!0.72$) and returns nearly identical values for the CTI-vs-generated and generated-vs-ground-truth columns (e.g., $0.72$ vs $0.73$), whereas RAGAS ($\approx\!0.70$ vs $\approx\!0.95$) and BERT-F1 ($\approx\!-0.15$ vs $\approx\!0.46$) diverge widely between the two [Table~\ref{tab:full_all}]. the same behavior appears in all four evaluation graphs [Fig.~\ref{fig:semantic_scorer_final_eval}], where our model traces nearly coincident trends (green line) across both comparisons. If the scorer captured surface lexical patterns rather than latent logic, the two comparisons ($1^{st}$ vs $2^{nd}$ and $3^{rd}$ vs $4^{th}$) would diverge, as they do for RAGAS and BERT-F1. Formally, if both CTI and rule are projected onto the same latent distribution, the CTI-to-generated and generated-to-ground-truth similarities should coincide, since the CTI-to-ground-truth score is unity $\big[\text{sim}\!\left(\mathcal{E}_c(\cti_i),\, \mathcal{E}_r(\idsrule_i)\right) = \text{sim}\!\left(\mathcal{E}_r(\idsrule_i),\, \mathcal{E}_r(\idsrule_i^{gt})\right)$, where $\text{sim}\!\left(\mathcal{E}_c(\cti_i),\, \mathcal{E}_r(\idsrule_i^{gt})\right) = 1\big]$. This consistency is strong evidence that the scorer represents logical relationships in latent space rather than matching lexical patterns, a crucial finding toward explainable, functionally meaningful evaluation of IDS rule generation.

    \textbf{(4) Self-reflection matters for complex rules (RQ-3).} Snort rules are comparatively simple, and LLMs often pass on the first shot, so the gain from one-shot (\textit{Single}) to validator-chained refinement (\textit{Chain}) is modest. YARA rules are more complex and frequently need 2--3 refinement iterations [Fig.~\ref{fig:e2e_semantic_scorer_final_eval}], producing large \textit{Single}$\to$\textit{Chain} increases in \textbf{Pass} [Table~\ref{tab:full_all}] (e.g., Mistral-Medium $597\!\to\!912$ and Qwen3-32B $648\!\to\!879$ on YARA, versus near-flat Snort). Aggregated over all matched configurations, refinement lifts the semantic score by $+18.0\%$ over one-shot [Table~\ref{tab:paired_improvement}], confirming that directed validator feedback yields measurable gains and supplies the generation trajectory LLMs lack for complex formats. By contrast, enabling agentic reasoning on the five reasoning-capable models contributes a negligible $\approx\!0\%$ on average [Table~\ref{tab:paired_improvement}], indicating that validator loop is the primary driver.

    \textbf{(5) Knowledge $\approx$ coverage; refinement $\approx$ quality.} Coverage scales with model size (observation 2) because larger models hold more of the threat- and format-specific knowledge needed to interpret a CTI and emit a valid rule. The \emph{quality} of accepted rules, by contrast, is size-agnostic: with self-reflective refinement, models of widely different sizes reach comparable Snort and YARA scores for a given CTI and its retrieved rules [Fig.~\ref{fig:e2e_semantic_scorer_final_eval}], as seen from Granite across its 3/8/30B variants. Scale therefore supplies the context knowledge that governs how many CTIs a model can translate, while the validator loop governs how good each accepted rule is.

    \textbf{(6) Domain fine-tuning substitutes scaling.} Cyber-tuned models confirm that coverage tracks domain knowledge rather than raw parameter count. Within their 7--8B class, the stronger ones match or beat same-size general models [Table~\ref{tab:full_all}]; more strikingly, on the simpler Snort format the domain-tuned Foundation-Sec (8B) reaches $797$ Pass, rivaling Mistral-Medium-128B ($800$) and surpassing far larger general models such as Qwen3-32B and Nemotron-49B, i.e., injected domain knowledge offsets a $16\times$ size gap. Scale's broader knowledge still prevails on the more complex YARA ($912$ for Mistral-Medium vs $\sim\!830$ for the domain-tuned 8B). Hence domain-alignment, can help limit scaling for bounded generation tasks.

\section{Conclusion}\label{section:conclusion}
    We presented \falcon, an agentic framework that transforms heterogeneous CTI into deployable Snort and YARA rules via retrieval, generation, and self-reflective syntactic, semantic, and performance validation. Its core is a novel ground-truth-free CTI-Rule semantic scorer that serves a dual retrieval-and-validation role. Across 18 LLMs we show that validator-driven self-reflection yields measurable quality gains ($+18\%$), largest for complex YARA rules, while agentic reasoning adds little, and that coverage is governed by domain knowledge. Achieving 84\% inter-rater agreement with SOC analysts, \falcon makes real-time, explainable, ground-truth-free IDS-rule automation feasible; future work targets proprietary CTI/rule corpora and SOAR integration for continuous, large-scale refinement.

\bibliographystyle{unsrt}
\bibliography{references}

\appendices

\section{List of Acronyms}\label{app:acronyms}
\begin{description}[font=\bfseries, style=sameline, leftmargin=5.4em, labelsep=0.6em, itemsep=1pt, parsep=0pt, topsep=2pt]
  \item[BERT] Bidirectional Encoder Representations from Transformers
  \item[BLEU] Bilingual Evaluation Understudy
  \item[BM25] Best Matching 25 (ranking function)
  \item[CFT] Contrastive Fine-Tuning
  \item[CTI] Cyber Threat Intelligence
  \item[DAG] Directed Acyclic Graph
  \item[FALCON] Feedback-driven ALignment framework for CTI-to-rule generatiON
  \item[Gen] Generated (rule)
  \item[GPT] Generative Pre-trained Transformer
  \item[GPU] Graphics Processing Unit
  \item[GT] Ground Truth
  \item[GUID] Globally Unique Identifier
  \item[HIDS] Host-based Intrusion Detection System
  \item[ICS] Industrial Control System
  \item[IDS] Intrusion Detection System
  \item[InfoNCE] Information Noise-Contrastive Estimation (loss)
  \item[IoC] Indicator of Compromise
  \item[KL] Kullback--Leibler (divergence)
  \item[LLM] Large Language Model
  \item[MAP] Mean Average Precision
  \item[MD5] Message-Digest Algorithm 5
  \item[NIDS] Network-based Intrusion Detection System
  \item[NT-Xent] Normalized Temperature-scaled Cross-Entropy (loss)
  \item[OSCTI] Open-Source Cyber Threat Intelligence
  \item[PCAP] Packet Capture
  \item[PE] Portable Executable
  \item[RAGAS] Retrieval-Augmented Generation Assessment
  \item[R@10] Recall at 10
  \item[ROUGE] Recall-Oriented Understudy for Gisting Evaluation
  \item[RQ] Research Question
  \item[SIEM] Security Information and Event Management
  \item[SME] Subject Matter Expert
  \item[SOAR] Security Orchestration, Automation, and Response
  \item[SOC] Security Operations Center
  \item[STIX] Structured Threat Information eXpression
  \item[TF-IDF] Term Frequency--Inverse Document Frequency
  \item[TTP] Tactics, Techniques, and Procedures
  \item[VAE] Variational Autoencoder
\end{description}

\section{Threats to Validity}\label{app:threats}
    \textbf{Construct.} \falcon uses the learned scorer $\Phi$ both as the semantic validator and as a headline metric, so any systematic bias in $\Phi$ propagates to retrieval and validation alike; we mitigate this through the CTI$\to$G vs GT$\to$G consistency analysis and independent SME ratings, but $\Phi$ is not an external ground truth, and \textbf{Pass} depends on the chosen validator thresholds. 
    
    \textbf{Internal.} Part of the \textit{Single}$\to$\textit{Chain} gain could arise from resampling rather than directed feedback; we do not fully separate the two, since each Chain run couples additional regenerations with the feedback signal.
    
    \textbf{External.} Evaluation covers open-source Snort and YARA under a standardized semi-structured CTI format and a fixed LLM suite. Hence, generalization to other engines (e.g., Suricata, Sigma), proprietary rule/CTI corpora, and noisier real-world CTI remains untested. 

    \textbf{Scope.} Consistent with our hypothesis (Section~\ref{section:problem_formulation}), we assume each CTI carries sufficient signature/behavior information for rule generation; adversarial crafted or information-poor CTI both violate this assumption, and \falcon's behavior in such cases is out of scope. The quality of the LLM-generated validator feedback is likewise orthogonal to this study and not separately analyzed.
    
    \textbf{Conclusion.} Most configurations are run once and the qualitative study rests on three SME raters (the authors) over 60 samples; broader, independent replication would strengthen the statistical claims.

\section{AI Disclosure}

    We used Claude and Gemini for refining experiment code, search and text refinement. These tools materially affected Section~\ref{section:experiment}.

\section{Acknowledgment}
    The authors acknowledge the use of Flaticon (\url{www.flaticon.com}) in paper's figures. Any opinions, findings, conclusions, or recommendations expressed in this material are those of the authors and do not necessarily reflect the views of their institution or the funding agencies.

\section{CTI-Rule Model Variant Statistics}

    This section presents the hyperparameter configurations (Table~\ref{tab:appendix_hyperparams}) and validation performance metrics for both the Snort (Table~\ref{tab:snort_hyperparam_runs}) and YARA (Table~\ref{tab:yara_hyperparam_runs}) semantic scorer variants evaluated across five optimization runs.

    \subsection{Hyperparameter Settings}
    
        \begin{table}[htbp]
        \caption{Hyperparameter Configurations for Tuning Runs}
        \label{tab:appendix_hyperparams}
        \centering
        \scriptsize
        \renewcommand{\arraystretch}{1.1}
        \begin{tabular}{c|ccccc}
        \hline
        \rowcolor{lightgray}
        \textbf{Run} & \textbf{Batch} & \textbf{Epochs} & \textbf{Learning Rate} & \textbf{LR Schedule} & \textbf{Temp.} \\ \hline\hline
        0 & 16  & 5  & $2 \times 10^{-5}$ & Constant      & 0.05 \\
        1 & 50  & 10 & $2 \times 10^{-5}$ & Constant      & 0.05 \\
        2 & 70  & 30 & $2 \times 10^{-5}$ & Constant      & 0.05 \\
        3 & 128 & 30 & $5 \times 10^{-5}$ & Warmup-Cosine & 0.05 \\
        4 & 70  & 50 & $2 \times 10^{-5}$ & Constant      & 0.07 \\ \hline
        \end{tabular}
        \end{table}

    \subsection{(Snort / YARA) CTI-Rule Model Variant Performance }

        \begin{table*}[htbp]
        \caption{Snort (NIDS) Semantic Scorer Model Evaluation across Different Runs}
        \label{tab:snort_hyperparam_runs}
        \centering
        \scriptsize
        \renewcommand{\arraystretch}{1.1}
        \setlength{\tabcolsep}{8pt}
        \begin{tabular}{ll|c|cccccc}
        \hline
        \rowcolor{lightgray}
        \textbf{Architecture} & \textbf{Backbone Encoder} & \textbf{Run} & \textbf{Recall@1} & \textbf{F1-Score} & \textbf{Threshold} & \textbf{Diag. Mean} & \textbf{Off-Diag. Mean} & \textbf{Val. Loss} \\ \hline\hline
        \multirow{15}{*}{Bi-Encoder} 
         & \multirow{5}{*}{\texttt{all-MiniLM-L6-v2}} 
           & 0 & 0.9551 & 0.9070 & 0.7012 & 0.9319 & 0.0825 & 0.0021 \\
         & & 1 & \cellcolor{softblue}0.9564 & 0.9164 & 0.6920 & 0.9278 & 0.0528 & 0.0138 \\
         & & 2 & 0.9539 & 0.9161 & 0.6974 & 0.9345 & 0.0550 & 0.0121 \\
         & & 3 & 0.9526 & 0.9302 & 0.7038 & 0.9452 & 0.0377 & 0.0200 \\
         & & 4 & 0.9551 & 0.9159 & 0.7038 & \cellcolor{softblue}0.9543 & 0.0052 & 0.0135 \\ \cline{2-9}
         & \multirow{5}{*}{\texttt{all-mpnet-base-v2}} 
           & 0 & 0.9539 & 0.9118 & 0.7036 & 0.9352 & 0.0594 & \cellcolor{softblue}0.0020 \\
         & & 1 & 0.9526 & 0.9310 & 0.7008 & 0.9283 & 0.0472 & 0.0134 \\
         & & 2 & 0.9551 & 0.9301 & 0.6966 & 0.9358 & 0.0510 & 0.0123 \\
         & & 3 & 0.9551 & 0.9360 & 0.6986 & 0.9423 & 0.0381 & 0.0202 \\
         & & 4 & 0.9551 & \cellcolor{softgreen}0.9443 & 0.7112 & \cellcolor{softgreen}0.9668 & 0.0073 & 0.0118 \\ \cline{2-9}
         & \multirow{5}{*}{\texttt{e5-base-v2}} 
           & 0 & 0.9526 & 0.9017 & 0.6909 & 0.9236 & 0.1215 & \cellcolor{softgreen}0.0019 \\
         & & 1 & 0.9551 & 0.9244 & 0.6960 & 0.9281 & 0.0744 & 0.0116 \\
         & & 2 & 0.9551 & 0.9292 & 0.6982 & 0.9251 & 0.0655 & 0.0114 \\
         & & 3 & \cellcolor{softblue}0.9564 & 0.9329 & 0.6951 & 0.9309 & 0.0491 & 0.0173 \\
         & & 4 & 0.9551 & 0.9324 & 0.7080 & 0.9532 & 0.0155 & 0.0123 \\ \hline\hline
        \multirow{15}{*}{Dual-Encoder} 
         & \multirow{5}{*}{\texttt{all-MiniLM-L6-v2}} 
           & 0 & 0.9539 & 0.8945 & 0.6762 & 0.8634 & 0.0394 & \cellcolor{softblue}0.0020 \\
         & & 1 & 0.9551 & 0.9132 & 0.6886 & 0.8966 & 0.0373 & 0.0129 \\
         & & 2 & \cellcolor{softgreen}0.9576 & 0.9276 & 0.6931 & 0.9105 & 0.0176 & 0.0113 \\
         & & 3 & 0.9551 & 0.9202 & 0.6902 & 0.9246 & 0.0408 & 0.0199 \\
         & & 4 & 0.9539 & 0.9211 & 0.7031 & 0.9504 & 0.0050 & 0.0134 \\ \cline{2-9}
         & \multirow{5}{*}{\texttt{all-mpnet-base-v2}} 
           & 0 & 0.9514 & 0.9151 & 0.6778 & 0.8442 & 0.0089 & \cellcolor{softblue}0.0020 \\
         & & 1 & \cellcolor{softblue}0.9564 & 0.9221 & 0.6777 & 0.8301 & \cellcolor{softblue}-0.0012 & 0.0125 \\
         & & 2 & 0.9539 & 0.9356 & 0.6818 & 0.8507 & \cellcolor{softgreen}-0.0019 & 0.0119 \\
         & & 3 & \cellcolor{softblue}0.9564 & 0.9299 & 0.6863 & 0.8859 & 0.0112 & 0.0213 \\
         & & 4 & \cellcolor{softblue}0.9564 & 0.9433 & 0.7022 & 0.9511 & 0.0012 & 0.0116 \\ \cline{2-9}
         & \multirow{5}{*}{\texttt{e5-base-v2}} 
           & 0 & 0.9489 & 0.9065 & 0.6707 & 0.8161 & 0.0253 & \cellcolor{softblue}0.0020 \\
         & & 1 & 0.9526 & 0.9122 & 0.6849 & 0.8765 & 0.0660 & 0.0123 \\
         & & 2 & 0.9526 & 0.9248 & 0.6838 & 0.8613 & 0.0308 & 0.0119 \\
         & & 3 & 0.9539 & 0.9351 & 0.6925 & 0.8980 & 0.0355 & 0.0181 \\
         & & 4 & 0.9551 & \cellcolor{softblue}0.9440 & 0.7001 & 0.9503 & 0.0068 & 0.0114 \\ \hline
        \end{tabular}
        \end{table*}
        
        \begin{table*}[htbp]
        \caption{YARA (HIDS) Semantic Scorer Model Evaluation across Different Runs}
        \label{tab:yara_hyperparam_runs}
        \centering
        \scriptsize
        \renewcommand{\arraystretch}{1.1}
        \setlength{\tabcolsep}{8pt}
        \begin{tabular}{ll|c|cccccc}
        \hline
        \rowcolor{lightgray}
        \textbf{Architecture} & \textbf{Backbone Encoder} & \textbf{Run} & \textbf{Recall@1} & \textbf{F1-Score} & \textbf{Threshold} & \textbf{Diag. Mean} & \textbf{Off-Diag. Mean} & \textbf{Val. Loss} \\ \hline\hline
        \multirow{15}{*}{Bi-Encoder} 
         & \multirow{5}{*}{\texttt{all-MiniLM-L6-v2}} 
           & 0 & 0.9476 & 0.9075 & 0.7075 & 0.9657 & 0.1632 & \cellcolor{softblue}0.0032 \\
         & & 1 & 0.9487 & 0.9262 & 0.7081 & 0.9649 & 0.0723 & 0.0073 \\
         & & 2 & \cellcolor{softblue}0.9498 & 0.9156 & 0.7089 & 0.9674 & 0.0953 & 0.0096 \\
         & & 3 & 0.9487 & 0.9347 & 0.7080 & 0.9672 & 0.0671 & 0.0130 \\
         & & 4 & \cellcolor{softgreen}0.9509 & 0.9355 & 0.7078 & \cellcolor{softgreen}0.9765 & 0.0067 & 0.0102 \\ \cline{2-9}
         & \multirow{5}{*}{\texttt{all-mpnet-base-v2}} 
           & 0 & \cellcolor{softgreen}0.9509 & 0.9335 & 0.7089 & 0.9632 & 0.1041 & 0.0056 \\
         & & 1 & \cellcolor{softgreen}0.9509 & 0.9300 & 0.7008 & 0.9628 & 0.0473 & 0.0101 \\
         & & 2 & \cellcolor{softgreen}0.9509 & 0.9394 & 0.7046 & 0.9559 & 0.0670 & 0.0111 \\
         & & 3 & \cellcolor{softgreen}0.9509 & 0.9326 & 0.7006 & 0.9470 & 0.0347 & 0.0136 \\
         & & 4 & \cellcolor{softgreen}0.9509 & \cellcolor{softgreen}0.9420 & 0.7107 & \cellcolor{softblue}0.9754 & 0.0106 & 0.0103 \\ \cline{2-9}
         & \multirow{5}{*}{\texttt{e5-base-v2}} 
           & 0 & \cellcolor{softblue}0.9498 & 0.9298 & 0.7009 & 0.9494 & 0.1193 & 0.0057 \\
         & & 1 & \cellcolor{softgreen}0.9509 & 0.9290 & 0.7026 & 0.9635 & 0.1401 & 0.0102 \\
         & & 2 & \cellcolor{softblue}0.9498 & 0.9314 & 0.7060 & 0.9645 & 0.1542 & 0.0107 \\
         & & 3 & \cellcolor{softblue}0.9498 & 0.9381 & 0.7059 & 0.9630 & 0.0780 & 0.0134 \\
         & & 4 & \cellcolor{softblue}0.9498 & 0.9335 & 0.7082 & 0.9746 & 0.0298 & 0.0110 \\ \hline\hline
        \multirow{15}{*}{Dual-Encoder} 
         & \multirow{5}{*}{\texttt{all-MiniLM-L6-v2}} 
           & 0 & 0.9487 & 0.9179 & 0.7038 & 0.9263 & 0.1520 & \cellcolor{softgreen}0.0025 \\
         & & 1 & 0.9476 & 0.9209 & 0.7041 & 0.9429 & 0.0878 & 0.0078 \\
         & & 2 & 0.9487 & 0.9184 & 0.7084 & 0.9577 & 0.1051 & 0.0095 \\
         & & 3 & \cellcolor{softblue}0.9498 & 0.9329 & 0.7039 & 0.9508 & 0.0639 & 0.0134 \\
         & & 4 & \cellcolor{softblue}0.9498 & 0.9313 & 0.7105 & 0.9692 & \cellcolor{softblue}0.0063 & 0.0104 \\ \cline{2-9}
         & \multirow{5}{*}{\texttt{all-mpnet-base-v2}} 
           & 0 & \cellcolor{softblue}0.9498 & 0.9214 & 0.6746 & 0.8456 & 0.0140 & 0.0039 \\
         & & 1 & \cellcolor{softgreen}0.9509 & 0.9392 & 0.6951 & 0.9041 & 0.0207 & 0.0104 \\
         & & 2 & \cellcolor{softgreen}0.9509 & 0.9351 & 0.6849 & 0.8838 & 0.0123 & 0.0102 \\
         & & 3 & \cellcolor{softgreen}0.9509 & 0.9336 & 0.6979 & 0.9185 & 0.0234 & 0.0128 \\
         & & 4 & \cellcolor{softgreen}0.9509 & \cellcolor{softblue}0.9412 & 0.7060 & 0.9625 & \cellcolor{softgreen}0.0046 & 0.0101 \\ \cline{2-9}
         & \multirow{5}{*}{\texttt{e5-base-v2}} 
           & 0 & \cellcolor{softblue}0.9498 & 0.9303 & 0.6896 & 0.8985 & 0.1064 & 0.0062 \\
         & & 1 & 0.9476 & 0.9103 & 0.6999 & 0.9321 & 0.1420 & 0.0109 \\
         & & 2 & 0.9487 & 0.9334 & 0.7015 & 0.9336 & 0.1204 & 0.0109 \\
         & & 3 & 0.9487 & 0.9345 & 0.7014 & 0.9390 & 0.0968 & 0.0138 \\
         & & 4 & \cellcolor{softgreen}0.9509 & 0.9397 & 0.7109 & 0.9666 & 0.0309 & 0.0113 \\ \hline
        \end{tabular}
        \end{table*}

\newpage
\section{Example End-to-End Rule Generation and Validation}\label{app:example_use_case}

This section provides a complete, worked-out example of the \falcon generative and validation pipeline using a YARA rule generation task. The progression demonstrates how heterogeneous CTI is contextualized against a retrieved rule, synthesized into a candidate rule using target-specific instructions, evaluated via serial validators, and refined using self-reflective feedback (answering RQ-1, RQ-2, and RQ-3).

\subsection{Input and Retrieval}
The generation process begins with unstructured CTI. This excerpt describes an evolved Mirai variant.

\begin{appbox}{Input Cyber Threat Intelligence ($\cti_i$)}
\textbf{Description:} Mirai botnet variant samples drop an ELF payload that prints the string ``getting busy here'' on launch and embeds the hardcoded C2 domain \texttt{cnc.mirai-evolution.io} in the \texttt{.rodata} section.... \{Trimmed for consistency\}
\end{appbox}

Using the semantic scorer, \falcon retrieves structurally linked rules from the deployed rule base to supply context, helping prevent redundant ``rule bloat''.

\subsection{Candidate Generation}
The \textit{Rule Generator LLM Agent} receives a rigid system prompt configuring its role, constraints, and target platform as extracted from the experiment code. 

\begin{appbox}[label={box:gen_instruction}]{YARA System Instruction ($\instruction$)}
\begin{lstlisting}[basicstyle=\ttfamily\scriptsize, breaklines=true, breakindent=0pt]
You are an expert malware-analysis engineer. Given a Cyber Threat Intelligence (CTI) description, produce ONE valid YARA rule that would detect the activity or artifacts described. Output ONLY the rule (rule keyword through closing brace), with no commentary, no markdown fences.
\end{lstlisting}
\end{appbox}

\begin{appbox}{One-Shot In-Context Example (YARA)}
\begin{lstlisting}[basicstyle=\ttfamily\scriptsize, breaklines=true, breakindent=0pt]
CTI:
Mirai botnet variant samples drop an ELF payload that prints the string 'getting busy here' on launch and embeds the hardcoded C2 domain 'cnc.mirai-evolution.io' in the .rodata section....

Rule:
rule MIRAI_Variant_Evolution {
    meta:
        description = "Mirai variant with hardcoded C2"
    strings:
        $s1 = "getting busy here"
        $s2 = "cnc.mirai-evolution.io"
    condition:
        uint32(0) == 0x464c457f and any of them
}
\end{lstlisting}
\end{appbox}

The LLM outputs an initial candidate rule. In this sample, the syntax is valid, but the \texttt{condition} block mistakenly drops the new C2 string reference constraint (\texttt{\$c2}).

\begin{appbox}{Initial Generated Candidate ($\idsrule^{(0)}_i$)}
\begin{lstlisting}[basicstyle=\ttfamily\scriptsize, breaklines=true]
rule MIRAI_Variant_Evolution {
    meta:
        description = "Mirai variant with C2"
    strings:
        $msg = "getting busy here"
        $c2 = "cnc.mirai-evolution.io"
    condition:
        uint32(0) == 0x464c457f and $msg
}
\end{lstlisting}
\end{appbox}

\subsection{Validation and Self-Reflection}
The candidate $\idsrule^{(0)}_i$ passes through short-circuit serial validation. Here, it passes syntax, but the semantic validator flags functional misalignment because the CTI semantic elements and the missing \texttt{\$c2} drop the score to $0.62$, which is below the passing threshold of $0.7107$. The chain-loop builds structured user feedback.

\begin{appbox}[label={box:semantic_fb}]{Validator Feedback ($\feedback^{(0)}_i$)}
\label{box:syntactic_fb}%
The YARA rule you just proposed failed validation at the \texttt{semantic} stage. Review the feedback and output ONLY a corrected YARA rule, no commentary.

\textbf{Syntactic check:}
\{"status": true, "error\_comment": null\}

\textbf{Semantic check} (similarity of the rule to the input CTI; higher is better):
\{"status": false, "scores": \{"sim": 0.62, "threshold": 0.7107\} \}
\end{appbox}

\subsection{Refined Output and Analyst Note}
Conditioned on this strict feedback, the LLM refines the rule condition to include the missing constraint. This second iteration yields a rule that fulfills the syntax, functional semantics (similarity $\ge 0.7107$ and surpassing $\epsilon_i^{\max}$), and performance validation.

\begin{appbox}{Revised Validated Rule ($\idsrule^{(1)}_i$)}
\begin{lstlisting}[basicstyle=\ttfamily\scriptsize, breaklines=true]
rule MIRAI_Variant_Evolution {
    meta:
        description = "Mirai variant with hardcoded C2"
    strings:
        $s1 = "getting busy here"
        $s2 = "cnc.mirai-evolution.io"
    condition:
        uint32(0) == 0x464c457f and all of them
}
\end{lstlisting}
\end{appbox}

On successful cycle completion, performance checks (complexity/latency constraints) are unified with semantic checks for final human sign-off.

\begin{appbox}[label={box:performance_fb}]{Analyst Note ($\feedback^f_i \cup \feedback^p_i$)}
\textbf{Semantic Alignment ($\epsilon$):} 0.94 (Exceeds 0.7107 threshold \& max retrieved context score)\\
\textbf{Performance Cost:} 12 (Passes $\le 20$ severity capability gate)\\
\textbf{Recommendation:} \texttt{UPDATE} -- Candidate logically subsumes \texttt{MIRAI\_Generic}. Recommend deployment and deprecation of retrieved legacy rule to mitigate overhead.
\end{appbox}

\onecolumn
\section{Dataset Description}
    This section presents the empirical distributions and profiles of the dataset used to train and evaluate the framework. Across the evaluation corpus, the dataset characteristics are analyzed along three central dimensions: the frequency of negative distractors paired with each CTI profile (Decoys per CTI''), the linguistic volume of the input threat intelligence (CTI length'') , and the syntactic footprint of the formal detection constraints (Rule length'') categorized by ground-truth (gold'') and outdated context (``decoy'') signatures. These structural properties are mapped distinctively across the 802 network-based (Snort) and 916 host-based (YARA)  threat profiles to ensure robust diversity across both detection mediums.

    \begin{figure*}[!ht]
        \centering \includegraphics[width=1\textwidth]{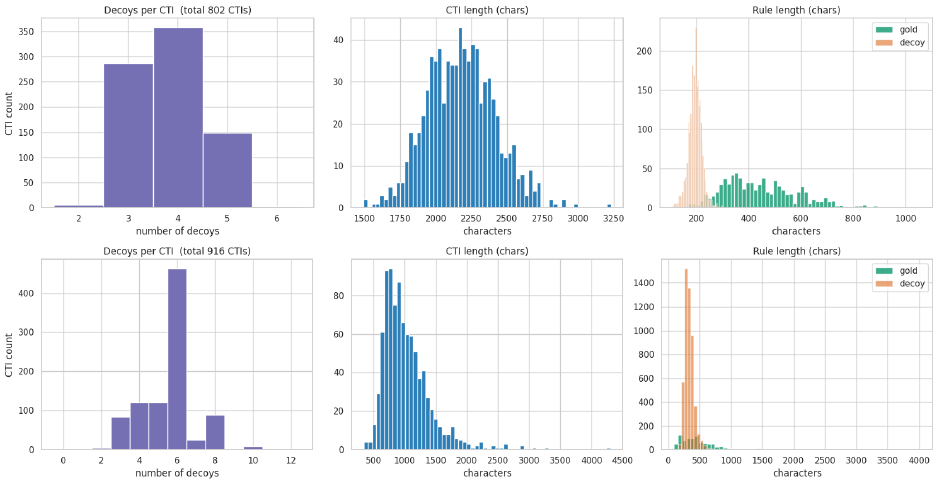}
        \caption{Distribution of the dataset for Snort (top row; 802 total CTIs) and YARA (bottom row; 916 total CTIs) platforms. The metrics trace the distribution of paired decoys per CTI (left) , the character length metrics of the source CTI text (middle) , and a comparative character length breakdown between ground-truth (gold) and legacy context (decoy) rulesets (right).}
        \label{fig:data_description_appendix}
    \end{figure*}

\section{Validator Ablation and Failure Mode Analysis}\label{app:validator_ablation}

    This section presents an ablation study analyzing 11,719 single-shot baseline failures and evaluates the corrective capacity of \falcon's feedback loop across 184 model runs. 
    
    Programmatic syntax flaws represent the primary barrier to automated rule deployment, accounting for 95\% (11,100 rules) of all single-shot rejections across the entire evaluation corpus. Specifically, Snort baseline failures are composed of 91\% syntax (1,899), 8\% semantic (169), and 1\% performance (13) errors. YARA baseline failures demonstrate an even higher syntactic fragility, skewing at 95\% syntax (9,201), 4\% semantic (341), and 1\% performance (96) rejections.
    
    Activating the validation chain effectively counteracts this structural fragility by translating execution errors into actionable adjustments. The loop successfully rescues 34.5\% of failing cases (717/2,081) for Snort and 51.5\% of failures (4,965/9,638) for complex YARA rule structures. Across all configurations, externalized feedback delivers an absolute mean lift of +0.130 (+18.0\%) to the final \texttt{pass\_all} capability gate over a mean convergence profile of 2.357 iterations, ensuring a 100\% win-rate over single-shot baselines. 
    
    Conversely, an ablation isolating test-time model reasoning (``thinking'') reveals a negligible impact on rule viability: averaged over all matched configurations [Table~\ref{tab:paired_improvement}], toggling thinking shifts the semantic score by only $-0.001$ ($-0.1\%$) over all-three-pass rules and $+0.006$ ($+0.7\%$) over syntax-pass rules. This confirms that compiler-driven external loop context---rather than internal token generation reasoning depth---is the definitive driver of deployable signature quality.

    \begin{figure*}[!ht]
        \centering \includegraphics[width=1\textwidth]{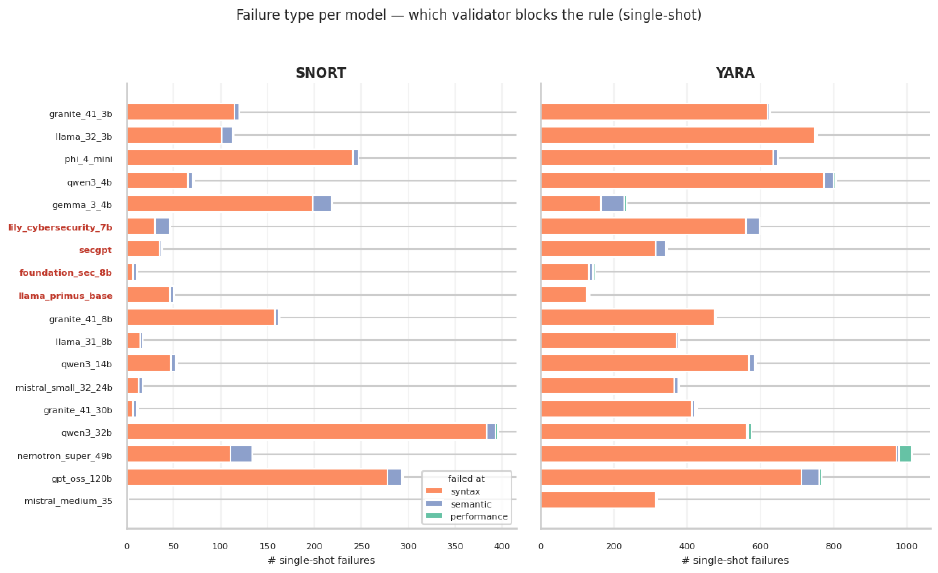}
        \caption{Comprehensive breakdown of single-shot baseline failure types across 18 evaluated language models for Snort (left) and YARA (right) platforms. The horizontal bars track baseline rejections by the specific validation component that blocked deployment, detailing the distribution of programmatic syntax flaws (orange), semantic functional mismatches (blue), and operational performance overheads (green).}
        \label{fig:validator_ablation_study_appendix}
    \end{figure*}

\newpage
\section{Cross-Modal Latent Space Consistency Analysis}\label{app:latent_consistency}

\subsection{Embedding Separability and Latent Space Alignment Shift}\label{app:embedding_separability}

\begin{figure*}[htbp]
    \centering
    \includegraphics[width=0.95\textwidth]{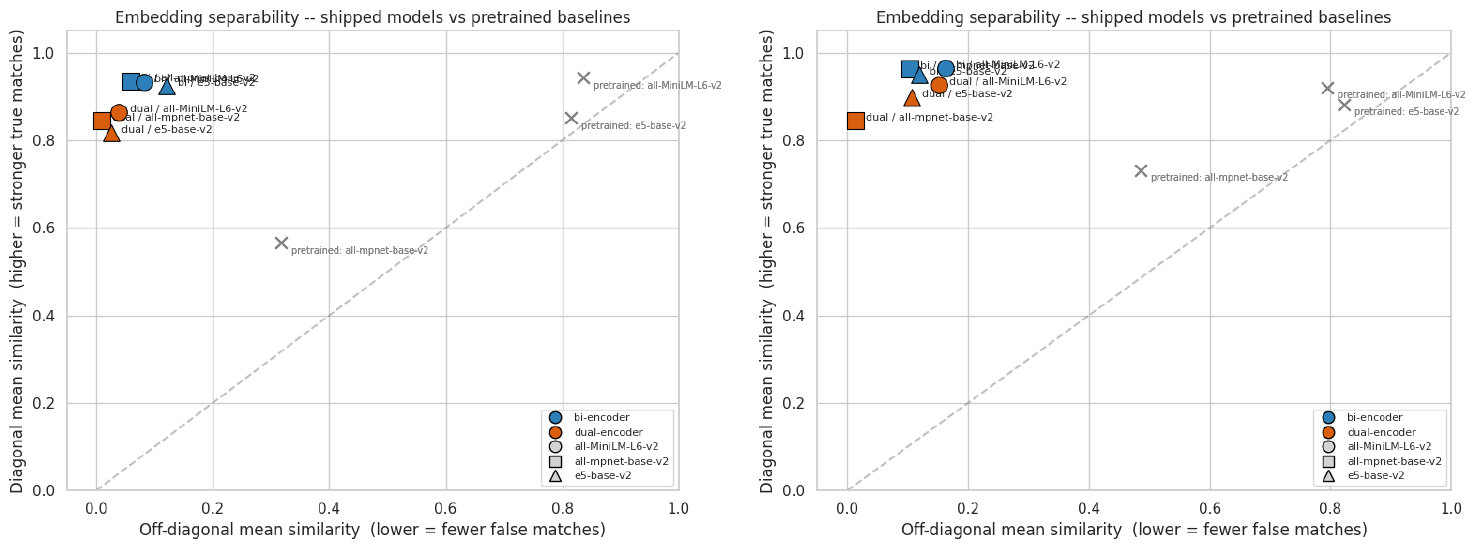}
    \caption{Embedding separability mapping out-of-the-box models against our for Snort (left) and YARA (right) in cross-modal spaces. The coordinate vector illustrates the migration from uncalibrated baselines (marked with 'X') to the optimized top-left quadrant. Maximizing vertical position ensures stronger true-match alignment, while minimizing horizontal position minimizes false-positive cross-modal matching.}
    \label{fig:separability_plots}
    \end{figure*}

To visualize how CFT restructures the cross-modal latent space, Figure~\ref{fig:separability_plots} maps our models against their pretrained foundation baselines, tracking true-match alignment (\textit{Diagonal Mean}) against false-match convergence (\textit{Off-Diagonal Mean}). Out-of-the-box baselines like \texttt{all-MiniLM-L6-v2} and \texttt{e5-base-v2} exhibit severe representation collapse in the upper-right quadrant, combining high diagonal similarity ($\approx 0.90$) with massive off-diagonal inflation ($\approx 0.80$). Similarly, pretrained \texttt{all-mpnet-base-v2} slightly reduces false matches ($\approx 0.32$) but suffers from depressed diagonal alignment ($\approx 0.56$). 

Contrastive fine-tuning drives a dramatic migration toward the ideal top-left corner, compressing off-diagonal false matches to a tight window ($0.01$ to $0.15$) while isolating true-match diagonal similarity between $0.85$ and $0.96$. Notably, Bi-Encoder variants (blue markers) demonstrate stronger true-match preservation over Dual-Encoder configurations (orange markers), confirming their superior capacity for context-aware retrieval and zero-day validation.
    
    \subsection{Cross-Modal Latent Space Consistency Validation}\label{app:latent_consistency_validation}

Here we present the trained semantic scorer as a ground-truth-free validator by comparing $\text{score}(\text{CTI}, \text{GenRule})$ against $\text{score}(\text{GTRule}, \text{GenRule})$. A reliable scorer's evaluation points should lie along the ideal $y=x$ diagonal line.
    
On the network-based \textbf{Snort} platform, the \texttt{CTI-Rule} scorer hugs the diagonal closely, achieving a Pearson correlation of $r = 0.888$ and a minimal mean absolute score gap of $|\Delta| = 0.023$. Baseline metrics reveal severe semantic distortions, with \texttt{BERT-F1} ($r = 0.428, |\Delta| = 0.355$) and \texttt{RAGAS} ($r = 0.511, |\Delta| = 0.269$) failing to capture consistent cross-modal alignment. 
    
For host-based \textbf{YARA} rules, our scorer demonstrates an even tighter calibration, reaching $r = 0.974$ and $|\Delta| = 0.012$. Lexical tracking via \texttt{BM25} correlates strongly on YARA ($r = 0.946, |\Delta| = 0.030$) due to identical structural identifier reuse, but collapses when evaluated against the abstract prose of Snort configurations ($r = 0.743, |\Delta| = 0.198$). This cross-paradigm stability confirms our regularized latent space measures true functional alignment rather than surface-level token matching.

    \begin{figure*}[!ht]
        \centering \includegraphics[width=1\textwidth]{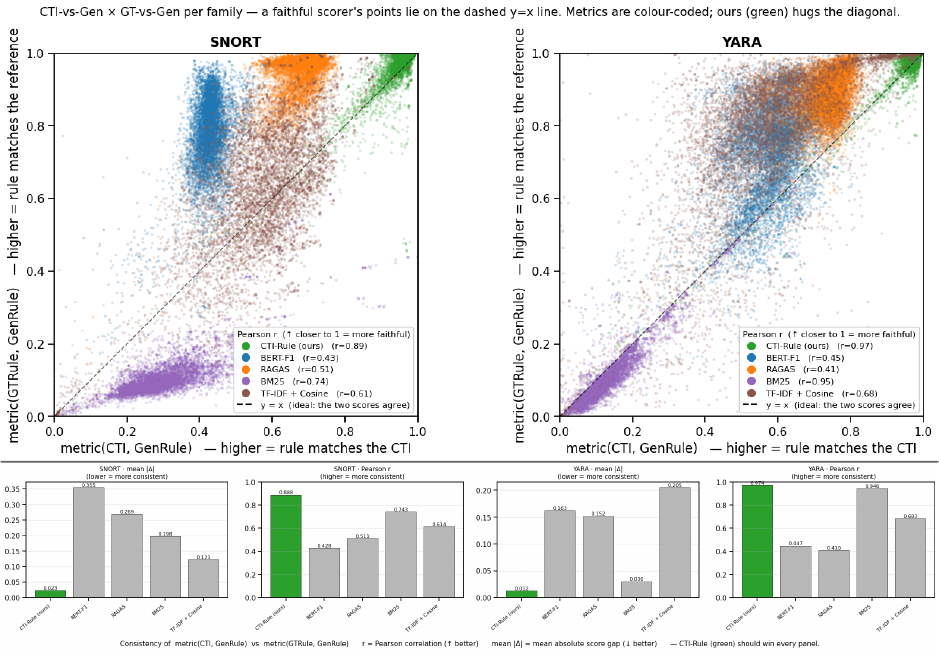}
        \caption{Cross-modal consistency evaluation mapping $\text{metric}(\text{CTI}, \text{GenRule})$ vs. $\text{metric}(\text{GTRule}, \text{GenRule})$ for Snort (left) and YARA (right) platforms. The scatter plots illustrate how closely each metric's proxy score conforms to the ideal $y=x$ diagonal agreement line. The lower bar charts quantify performance via the mean absolute score gap ($|\Delta|$, lower is better) and the Pearson correlation coefficient ($r$, higher is better).}
        \label{fig:semantic_scorer_eval}
    \end{figure*}
\end{document}